\documentclass[%
 reprint,
 amsmath,amssymb,
 aps,
prx,
superscriptaddress,
longbibliography
]{revtex4-2}

\usepackage{graphicx}
\usepackage{dcolumn}
\usepackage{bm}
\usepackage{bbold}
\usepackage{amsmath}

\newcommand{\id}{\mathbb{1}} 

\begin{document}

\preprint{APS/123-QED}

\title{On-demand coherent perfect absorption in complex scattering systems: \\time delay divergence and enhanced sensitivity to perturbations}

\author{Philipp del Hougne}
\affiliation{Univ Rennes, INSA Rennes, CNRS, Institut d'Electronique et des Technologies du num\'{e}Rique, UMR–6164, F-35000 Rennes, France}

\author{K. Brahima Yeo}
\affiliation{Univ Rennes, INSA Rennes, CNRS, Institut d'Electronique et des Technologies du num\'{e}Rique, UMR–6164, F-35000 Rennes, France}

\author{Philippe Besnier}
\affiliation{Univ Rennes, INSA Rennes, CNRS, Institut d'Electronique et des Technologies du num\'{e}Rique, UMR–6164, F-35000 Rennes, France}

\author{Matthieu Davy}
\affiliation{Univ Rennes, INSA Rennes, CNRS, Institut d'Electronique et des Technologies du num\'{e}Rique, UMR–6164, F-35000 Rennes, France}

\date{\today}

\begin{abstract}

Non-Hermitian photonic systems capable of perfectly absorbing incident radiation recently attracted much attention both because fundamentally they correspond to an exotic scattering phenomenon (a real-valued scattering matrix zero) and because their extreme sensitivity holds great technological promise. The sharp reflection dip is a hallmark feature underlying many envisioned applications in precision sensing, secure communication and wave filtering. However, a rigorous link between the underlying scattering anomaly and the sensitivity of the system to a perturbation is still missing. Here, we develop a theoretical description in complex scattering systems which quantitatively explains the shape of the reflection dip. We further demonstrate that coherent perfect absorption (CPA) is associated with a phase singularity and we relate the sign of the diverging time delay to the mismatch between excitation rate and intrinsic decay rate. We confirm our theoretical predictions in experiments based on a three-dimensional chaotic cavity excited by eight channels. Rather than relying on operation frequency and attenuation inside the system to be two free parameters, we achieve ``on-demand'' CPA at an arbitrary frequency by tweaking the chaotic cavity's scattering properties with programmable meta-atom inclusions. Finally, we theoretically prove and experimentally verify the optimal sensitivity of the CPA condition to minute perturbations of the system.

\end{abstract}

\maketitle

\section{Introduction}

The scattering of waves as they interact with matter is the basis of countless experimental methods, a prominent example being imaging. Recently, many exotic scattering phenomena such as perfect absorption, exceptional points or bound states in the continuum have been extensively studied due to their disruptive potential in areas such as sensing and computing; fundamentally, they can be understood in terms of analytical properties of the associated scattering matrix~\cite{krasnok2019anomalies}. Indeed, any wave scattering process is fully characterized by the distribution of poles and zeros of the scattering matrix in the complex frequency plane~\cite{grigoriev2013optimization,krasnok2019anomalies}. Poles and zeros are spectral singularities associated with outgoing and incoming boundary conditions, respectively. By including the appropriate amount of gain in the system, a pole can be pulled up onto the real frequency axis: the scattering matrix will have an infinite eigenvalue and lasing occurs. Conversely, by including the appropriate amount of loss, a zero can be pulled down onto the real frequency axis, resulting in a zero eigenvalue of the scattering matrix and coherent perfect absorption (CPA)~\cite{chong2010coherent,Baranov2017}. Incident radiation corresponding to the eigenvector associated with the zero eigenvalue will be perfectly absorbed. CPA can be interpreted as a generalization of the critical coupling condition~\cite{cai2000observation} and can be understood as the time reverse operation of a laser (anti-laser)~\cite{wan2011time,wong2016lasing}. The generality of these concepts implies that they also apply to random scattering media. Indeed, a “random laser” resonantly enhances light by multiple scattering inside a disordered medium~\cite{cao2003lasing,wiersma2008physics}. Recently, the feasibility of realizing CPA in random scattering media and chaotic cavities has been studied theoretically and demonstrated experimentally~\cite{fyodorov2017distribution,li2017random,Pichler2019,chen2020perfect}, overcoming the immense difficulty of balancing excitation and decay rate of a random system.

A hallmark signature of a system with a scattering matrix zero on the real frequency axis is a very pronounced dip of the energy reflected off the system as a function of frequency or any other local or global system parameter, evidencing an extreme sensitivity to tiny perturbations. This feature is at the heart of the concept's technological relevance: for regular systems, it was, for instance, leveraged to demonstrate coherent modulation of light with light, i.e.~without any non-linearity~\cite{zhang2012controlling,bruck2013plasmonic,rao2014coherent}; for randomly scattering systems, the extreme sensitivity is the basic ingredient of a recently demonstrated physically secure wireless communication scheme~\cite{del2020PA} as well as of envisioned precision-sensing applications~\cite{chen2020perfect,del2020PA}. Intuitively, one may explain this extreme sensitivity with the fact that the incident radiation is trapped for an infinite time~\cite{chong2010coherent} when a real-valued scattering matrix zero is accessed: the longer the wave's lifetime inside a chaotic system, the more likely it is that its evolution is impacted even by a tiny perturbation. Nonetheless, a rigorous explanation of the physical origins of this extreme sensitivity and its link to the underlying scattering anomaly is to date missing. 

In our work, we fill this gap by studying the time delay of waves at the CPA condition inside a random medium. Our theoretical model quantitatively explains the shape of the reflection dip and relates the sign of the diverging time delay to the difference between the system’s excitation and decay rate. We further analytically demonstrate the CPA condition’s optimal sensitivity to minute perturbations, irrespective of their location inside a chaotic system. Our theoretical findings are corroborated with random matrix simulations as well as experiments in the microwave domain involving a chaotic cavity. To facilitate accessing a real-valued zero experimentally, we tune our disordered system's scattering properties with programmable meta-atom inclusions to a state in which a zero of the scattering matrix hits the horizontal axis~\cite{del2020PA}. This procedure enables the ``on-demand'' observation of a CPA condition at any desired frequency and without reliance on attenuation being dominated by a localized and tunable loss center. Our results pave the way for sensors with optimal sensitivity to minute perturbations of disordered matter such as tiny intrusions, defaults, changes in temperature or concentration.

\section{Theoretical Model of Time Delays in Random CPA}

\subsection{Dip in Frequency-Dependent Reflection Coefficient}
The wave-matter interaction in a complex scattering system is fully characterized by its scattering matrix $S(\omega)$ which relates an incoming field $\psi_{in}$ to the corresponding outgoing field $\psi_{out}$ via $\psi_{out} = S(\omega) \psi_{in}$.
In a system without any absorption or loss, $S(\omega)$ is unitary and its zeros ($z_m=\omega_m+i\Gamma_m/2$) and poles ($\Tilde{\omega}_m = \omega_m - i\Gamma_m/2$) are symmetrically placed in the upper and lower half of the complex frequency plane, respectively. 
$\omega_m$ and $\Gamma_m$ denote central frequency and linewidth of the system's $m$th resonance. 
In non-Hermitian systems, the presence of attenuation (or gain) $\Gamma_a$ moves the zeros in the complex plane: $z_m =\omega_m +i(\Gamma_m - \Gamma_a)/2$. 
When attenuation losses exactly balance dissipation through the channels for a given zero (labelled with the subscript $n$ in the following), that is $\Gamma_a = \Gamma_n$, the zero crosses the real frequency axis such that $z_n = \omega_n$. Then, $S(\omega_n)$ has a zero eigenvalue such that the corresponding eigenvector $\psi_{\text{CPA}}$ satisfies $S(\omega_n)\psi_{\text{CPA}} = 0$ and the multi-channel reflection coefficient $R(\omega_n) = \|\psi_{out}\|^2$ vanishes: $R(\omega_n) \rightarrow 0$.

The vanishing reflection at the CPA condition suggests that the wave is infinitely trapped in the medium which we expect to translate into a diverging time delay. To rigorously investigate this hallmark property for CPA in a complex scattering system, we avail ourselves of a non-perturbative effective Hamiltonian formalism. 
Therein, a $M\times M$ Hamiltonian $H_0$ describes the internal system, its coupling to the $N$ channels is characterized by a matrix $V$, and $S(\omega) = \id - iV^T [\omega\id - H_0 + i(VV^T + \Gamma_a \id)/2]^{-1} V$ \cite{Rotter2001pre,Fyodorov2005,Rotter2009Jphys}. The scattering matrix can also be decomposed in terms of the system's natural resonances (the poles) as \cite{Rotter2001pre,Alpeggiani2017,Davy2019}

\begin{equation}
    S(\omega) = \id - i \Sigma_{m=1}^M \frac{W_mW_m^T}{\omega-\omega_m +i(\Gamma_m + \Gamma_a)/2}.
\label{eq:decomp_S_resonances}    
\end{equation}

\noindent The complex eigenfrequencies $\Tilde{\omega}_m = \omega_m - i\Gamma_m/2$ are the eigenvalues of the effective Hamiltonian $H_{\text{eff}} = H_0-iVV^T/2$ and the modal vectors $W_m$ are the projections of the eigenfunctions $\phi_m$ of $H_{\text{eff}}$ onto the channels: $W_m = V^T\phi_m$. Using the completeness of the eigenfunctions, it can be demonstrated (see SM) that the eigenstate corresponding to a CPA condition at $\omega =\omega_n$, $S(\omega_n)\psi_{\text{CPA}} =0$, is the time-reverse of the modal wavefront: $\psi_{\text{CPA}} = W_n^* / \| W_n \|$. $\psi_{\text{CPA}}$ hence provides maximal excitation of the selected mode~\cite{Davy2019,del2020experimental}.
The interpretation of $W_n^*$ as the time-reversed output of a lasing mode (if loss mechanisms were replaced by gain mechanisms of equal strength) led to the term ``anti-laser'' \cite{chong2010coherent,wong2016lasing,Pichler2019}, an analogy that should be used with caution since it neglects essential nonlinear processes in laser operation.

Any realistic experimental observation of CPA is however inevitably confronted with multiple practical imperfections: (i) noise corrupts the measurement of $S$ such that one measures $S+\Delta S$, (ii) a small mismatch $\Delta\Gamma$ between the modal linewidth and losses: $\Gamma_a  = \Gamma_n +\Delta\Gamma$, (iii) a frequency shift $\Delta\omega = \omega - \omega_n$. 
In the vicinity of the CPA condition, the resonance associated with the real-valued zero dominates the sum in Eq.~(1). The latter implies that $\psi_{\text{CPA}}$ is still an eigenvector of $S$ with a reflection coefficient which is therefore not zero but (see SM for algebraic details)

\begin{equation}
    R(\omega) = \frac{4 \Delta \omega ^2 + (\Delta\Gamma)^2}{4 \Delta \omega ^2 + (2\Gamma_n+\Delta\Gamma)^2} + \frac{\|\Delta S\|_F^2}{N}.
\label{eq:R}    
\end{equation}

\noindent We will show below that this equation explains the characteristic shape of the reflection dip and can be exploited to extract the experimental parameters by fitting the measured data with this model.  

\subsection{Time Delay}

Having an analytical description of the reflection dip at the CPA condition in a random system, we can move on to study the associated time delays. 
The time delay of an incoming wavefront scattered in a multichannel system is determined via the Wigner-Smith operator $Q(\omega) = -i S(\omega)^\dagger \partial_\omega S(\omega)$ which involves the derivative of $S(\omega)$ with angular frequency: 

\begin{equation}\label{eq_tau}
    \tau(\omega) = \frac{\psi_{\text{in}}^\dagger Q(\omega) \psi_{\text{in}}}{\psi_{\text{in}}^\dagger S(\omega)^\dagger S(\omega) \psi_{\text{in}}}.     
\end{equation}
 
\noindent The real part of the complex-valued $\tau(\omega)$ is related to the frequency derivative of the scattering phase and can hence be interpreted as the delay of reflected intensity for an incoming pulse with vanishing bandwidth. The imaginary part of $\tau(\omega)$ is related to the variation of reflected intensity with frequency~\cite{Bohm2018,fan2005principal,Durand2019}. 
\noindent Our model also allows us to estimate the real and imaginary parts of $\tau(\omega)$ at the CPA condition, $\psi_{\text{in}} = \psi_{\text{CPA}}$, (see SM): 

\begin{equation}
    \text{Re}[\tau(\omega)] = \frac{1}{R(\omega)} \frac{4\Gamma_n (4\Delta\omega^2 - \Delta\Gamma (2\Gamma_n + \Delta\Gamma))}{[4\Delta\omega^2 + (2\Gamma_n + \Delta\Gamma)^2]^2}
\label{eq:tau_r}    
\end{equation}

\noindent and

\begin{equation}
    \text{Im}[\tau] = -\frac{4\Gamma_n}{4\Delta\omega^2 + \Delta\Gamma^2} \frac{4\Delta\omega (\Gamma_n + \Delta\Gamma) }{4\Delta\omega^2 + (2\Gamma_n + \Delta\Gamma)^2},
\label{eq:tau_i}    
\end{equation}

\noindent where $R(\omega)$ is given by Eq.~(\ref{eq:R}). Obviously both the real and imaginary parts of $\tau(\omega)$ diverge as $\Delta \omega \rightarrow 0$ and $\Delta \Gamma \rightarrow 0$, which confirms the intuition that the wave injected into the system is trapped for an infinitely long time at the CPA condition. 

Surprisingly, we observe a phase transition of $\text{Re}[\tau(\omega = \omega_n)]$ as the amount of losses increases and surpasses the modal linewidth. 
When the zero $z_n = \omega_n +i(\Gamma_n - \Gamma_a)/2$ is located in the upper half of the complex frequency plane ($\Gamma_a < \Gamma_n$), the time delay $\text{Re}[\tau(\omega)]$ is positive. At the crossover of the real frequency axis ($\Gamma_a = \Gamma_n$), a singularity occurs and $|\text{Re}[\tau(\omega)]|$ diverges. When losses dominate the coupling to channels, $\Gamma_a > \Gamma_n$, $\text{Re}[\tau(\omega)]$ becomes negative. We note here that negative time delays have previously been found for wavefronts that are strongly absorbed by lossy resonant targets within multiply scattering media~\cite{muga1998negative,tanaka2003propagation,Durand2019}. Negative time delays arise due to the distortion of the incident pulse for which the intensity is more strongly absorbed at long lifetimes than at early times (see SM).

\subsection{Sensitivity to Perturbations}

We now seek to demonstrate that the strong enhancement of the delay time provides an extreme sensitivity of the outgoing field to tiny perturbations within the system. To establish this link, let us begin by considering the \textit{generalized} Wigner-Smith operator $Q_\alpha = -iS^{\dagger} \partial_\alpha S$~\cite{Ambichl2017b} defined with respect to a change of a parameter $\alpha$ of the system; $\alpha$ can be a local or a global parameter. In systems where $S$ is close to unitarity, the optimal eigenstates of $Q_\alpha$ provide the optimal wavefronts to locally manipulate a perturber \cite{Ambichl2017b,Horodynski2019}. In analogy with the delay time in Eq.~(\ref{eq_tau}), the variations of the outgoing field for a change of $\alpha$ are encapuslated within the complex parameter $\tau_\alpha = \psi_{\text{in}}^\dagger Q_\alpha \psi_{\text{in}} / R$. The relation between the delay time $\tau(\omega)$ and $\tau_\alpha$ at the CPA condition can be established using the modification of the system's resonances due to the perturbation. 

First, we note that upon injection of the CPA wavefront ($\psi_{\text{in}} = \psi_{\text{CPA}}$), the variation of the outgoing field in the vicinity of the CPA condition results from the change of a single eigenstate (labelled $n$) in Eq.~(1). This property yields a linear relation between the projection of the CPA wavefront on $Q_\alpha$ and on the \textit{generalized} Wigner-Smith operator applied to a perturbation of $\omega_n$, $Q_{\omega_n} = -i S^\dagger \partial_{\omega_n} S$, 

\begin{equation}
    \psi^\dagger_{\text{CPA}} Q_\alpha \psi_{\text{CPA}} = [\partial_\alpha \omega_n] \psi^\dagger_{\text{CPA}} Q_{\omega_n} \psi_{\text{CPA}}.
\end{equation}

\noindent The perturbation-induced shift in $\omega_n$ is given by $\partial_\alpha \omega_n  = \beta \nabla |\phi(r)|^2$ for local perturbations, where $\nabla |\phi(r)|^2$ is the gradient of the energy density taken in the direction of the displacement and $\beta$ depends on the geometry of the perturber \cite{Barth1999}. 

Second, in the approximation of a single resonance contribution, the projection on the operators $Q = -i S^\dagger \partial_{\omega} S$ and $Q_{\omega_n}$ yields the same result except for a global minus sign. This provides the sought-after relation between $\tau(\omega)$ and $\tau_\alpha$ for the CPA condition:

\begin{equation}
    \tau_\alpha= - [\partial_\alpha \omega_n]  \tau(\omega).
\label{eq:tau_alpha}
\end{equation}

\noindent The divergence of $\tau(\omega)$ hence leads to an extreme sensitivity of the outgoing field. In our case, the CPA wavefront does not provide focusing on a perturber but is the optimal wavefront to detect a tiny variation anywhere within the cavity. We emphasize that $\psi_{\text{CPA}}$ is also the optimal eigenvector of the operator $-iS^{-1} \partial_\alpha S$ (see SM). This property results from the vanishing eigenvalue of $S$ leading to a pseudo-inverse matrix $S^{-1}$ dominated by a single eigenstate.

Equation~(\ref{eq:tau_alpha}) directly provides the sensitivity of the reflection coefficient at the CPA condition. Using that $\partial_\alpha R = -2 \text{Im}[\psi_{\text{CPA}}^\dagger Q_\alpha \psi_{\text{CPA}}]$, we obtain

\begin{equation}
    \frac{1}{R}\partial_\alpha R=  [\partial_\alpha \omega_n]  \text{Im}[\tau(\omega)].
\label{eq:sensitivity}
\end{equation}

\noindent The derivative of the logarithm of the reflection coefficient ($\partial_\alpha \text{log}(R) = [\partial_\alpha R]/R$) hence increases extremely rapidly in the vicinity of the CPA condition. This unique feature makes it possible to finely characterize the strength of the perturbation from the shape of the reflection dip. Because the change in $R$ is proportional to the change in $\omega_n$, this dip is anew given by Eq.~(\ref{eq:R}) in which $\Delta\omega$ has to be replaced with $\Delta\omega_n \sim \Delta \alpha [\partial_\alpha \omega_n] = \Delta \alpha [\beta \nabla |\phi(r)|^2]$.

We note that another slightly different operator has been introduced in recent related work to maximize the measurement precision of an observable parameter~\cite{Bouchet2020}. The eigenstates of the operator $F_\alpha = (\partial_\alpha S)^\dagger \partial_\alpha S$ have indeed been identified as ``maximum information states'' maximizing the Fisher-information related to the parameter $\alpha$. $F_\alpha$ coincides with $Q_\alpha$ only for a unitary scattering matrix. The results in Ref.~\cite{Bouchet2020} however differ sharply from our present work in two important ways: first, while Ref.~\cite{Bouchet2020} considers a ``random'' configuration of a disordered medium, we operate under the very special CPA condition for which $S$ has a real-valued zero; second, while Ref.~\cite{Bouchet2020} identifies a so-called ``maximum information state'' that is specific to the observable of interest (e.g. location of the pertuber), our CPA condition yields an optimal sensitivity to \textit{any} perturbation irrespective of its location.

\begin{figure*}
    \centering
    \includegraphics[width=16cm]{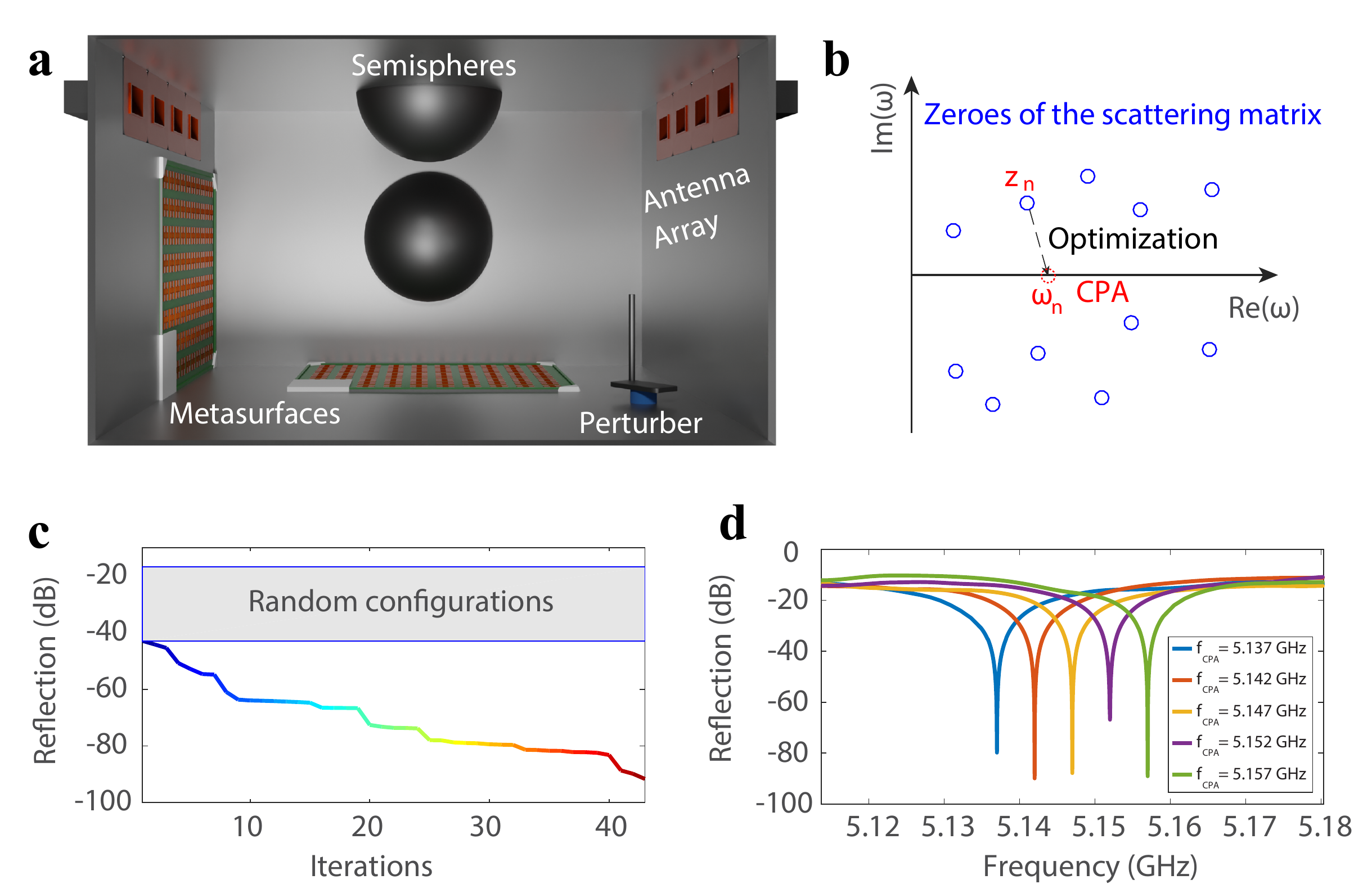}
    \caption{\textbf{``On-demand'' realization of CPA in a programmable complex scattering enclosure.} \textbf{a,} Experimental setup consisting of a three-dimensional electrically large irregular metallic enclosure equipped with two arrays of 1-bit reflection-programmable meta-atoms to tune the system's scattering properties. The system is connected to eight channels via waveguide-to-coax transitions. Small perturbations of the system can be induced by rotating a metallic rod placed on a metallic platform. 
    \textbf{b,} Illustration of operation principle in the complex frequency plane. By tuning the system's scattering properties with the programmable meta-atoms, a zero of the scattering matrix is moved onto the real frequency axis at a target horizontal position (here 5.147~GHz).
    \textbf{c,} Dynamics of an example iterative optimization of the meta-atom configurations. 
    \textbf{d,} Spectrum of the multi-channel reflection coefficient $R(\omega)$ for five optimized systems targeting five distinct regularly spaced nearby target frequencies between 5.137~GHz and 5.157~GHz.}
    \label{fig:setup}
\end{figure*}

\begin{figure*}
    \centering
    \includegraphics[width=18cm]{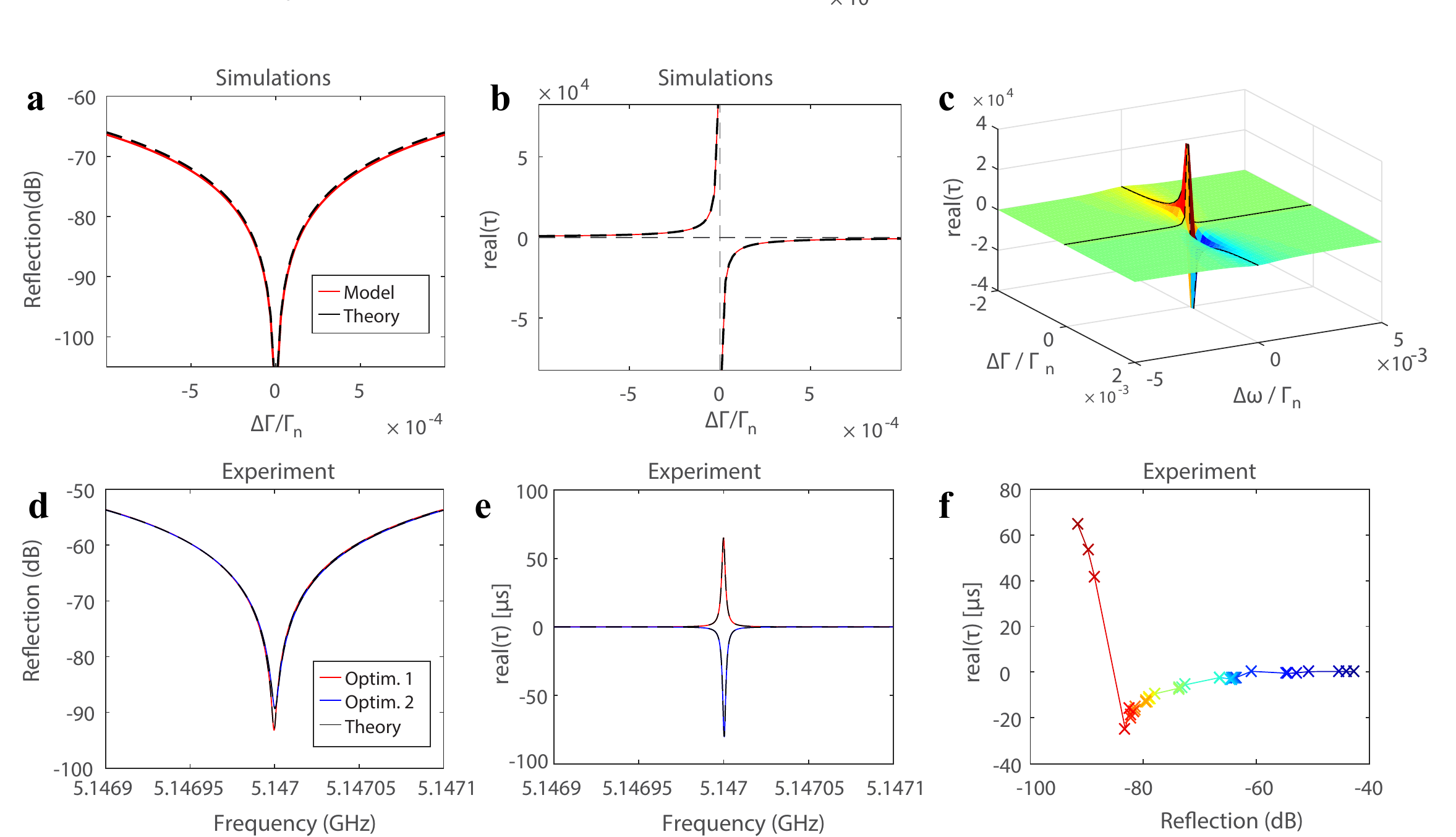}
    \caption{\textbf{CPA signature on time delays.} Multichannel reflection coefficient $R(\omega)$ (\textbf{a}) and time delay $\text{Re}[\tau(\omega)]$ (\textbf{b}) found in random matrix simulations with an effective Hamiltonian model (red lines) in the vicinity of a CPA condition at $\omega_n$ for a mode with linewidth $\Gamma_n$, plotted as a function of linewidth detuning $\Delta \Gamma / \Gamma_n = (\Gamma_a - \Gamma_n)/\Gamma_n$. The time delay has been normalized by its value in absence of absorption. These numerical results are in excellent agreement with our theoretical predictions (black dashed lines) given by Eq.~(\ref{eq:R}) and Eq.~(\ref{eq:tau_r}). 
    \textbf{c,} Variations of the time delay with $\Delta \Gamma / \Gamma_n$ and frequency detuning $\Delta\omega/\Gamma_n = (\omega - \omega_n)/\Gamma_n$ are visualized as three-dimensional surface. 
    The experimental results in \textbf{d} and \textbf{e} as a function of frequency detuning are also perfectly fitted with our theory for two different realizations of the CPA condition. 
    \textbf{f,} Evolution of time delay during an example optimization. The absolute time delay $|\text{Re}[\tau(\omega)]|$ increases as $R(\omega_0)$ is minimized but the sign of $\text{Re}[\tau(\omega)]$ jumps from negative to positive after 40 iterations. This is a signature of the time-delay singularity. The color-code indicating the iteration index is the same as in Fig.~1c.}
    \label{fig:time-delay}
\end{figure*}

\section{Experimental Measurement of Time Delays}

Having established a theoretical model for the characteristic reflection dip and delay time associated with CPA, we now seek to verify its validity in experiments. Experimentally realizing CPA in a random medium is a very challenging task that was only mastered recently for the first time~\cite{Pichler2019,chen2020perfect}. These early realizations relied on both $\omega$ and $\Gamma_a$ being freely tunable parameters to identify a setting in which one zero of $S(\omega)$ lies on the real frequency axis. For our experiments, we consider a more realistic three-dimensional complex scattering enclosure with fixed homogeneously distributed losses and we fix the working frequency to $5.147$~GHz. In order to realize CPA ``on demand'' without control over $\omega$ and $\Gamma_a$, we dope our system with reflection-programmable meta-atoms (see Appendix).  In our experiment illustrated in Fig.~\ref{fig:setup}a and as detailed in the Appendix, $N=8$ channels are connected to a chaotic cavity and an iterative algorithm is used to optimize the configuration of the 304 meta-atoms. The latter allow us to tweak $S$ such that one of its zeros hits the real frequency axis~\cite{del2020PA}, as illustrated schematically in Fig.~\ref{fig:setup}b. Moreover, our setup includes an irregular metallic structure attached to a rotation stage; we can thus identify different realizations of CPA by rotating this ``mode-stirrer'' to different positions and optimizing the meta-atom configurations for each position.

Random configurations of the meta-atoms yield reflection values between $-20\ \mathrm{dB}$ and $-40\ \mathrm{dB}$. Starting from a configuration corresponding to roughly $-40\ \mathrm{dB}$, our optimization algorithm which minimizes the reflection $R = \|S\psi_{in}\|^2 $ eventually identifies a setting with $R\sim 4.85 \times 10^{-10} = -93$~dB, as depicted in Fig.~\ref{fig:setup}c. We then repeat the experiment for other predefined frequencies. In each case, the corresponding reflection spectrum in Fig.~\ref{fig:setup}d displays the expected very narrow and deep dip at the desired working frequency.

We can now test our theoretical model's prediction for the reflection and delay time at the CPA condition. Before confronting it with our experimental data, we perform a random matrix simulation for which the internal Hamiltonian $H_0$ is a a real symmetric matrix drawn from the Gaussian orthogonal ensemble and $V$ is a real random matrix with Gaussian distribution~\cite{kuhl2013microwave}. We select a resonance and explore the variations of $R(\omega)$ and $\text{Re}[\tau(\omega)]$ near the CPA condition as a function the normalized linewidth mismatch $\Delta\Gamma / \Gamma_n$ and frequency detuning $\Delta\omega / \Gamma_n$. As shown in Fig.~\ref{fig:time-delay}a-c, the dip in the reflection coefficient and the singularity of the time delay found in simulations are perfectly reproduced by Eqs.~(\ref{eq:R}) and (\ref{eq:tau_r}). In particular, Fig.~\ref{fig:time-delay}b highlights the divergence of $\text{Re}[\tau(\omega)]$ as $\Delta\Gamma \rightarrow 0$ with positive (negative) values when the zero lies in the upper (lower) complex plane.

To compare our theoretical model with our experiment, we extract the parameters involved in the model, that is $\Gamma_n$, $\Delta\Gamma$ and $\|\Delta S\|_F^2$, by fitting our model to the spectra of $R(\omega)$ and $\text{Re}[\tau(\omega)]$. We thereby obtain a linewidth $\gamma_{\text{CPA}} = \Gamma_{\text{CPA}}/(2\pi) \sim 51$~MHz, a mismatch $\Delta\Gamma/(2\pi) \sim 1.3$~kHz and a relative noise level $\|\Delta S \|_F^2 / \|S \|_F^2 \sim 3\times10^{-9} $. As seen in Fig.~\ref{fig:time-delay}d,e, our model exactly fits our experimental data on a frequency range smaller than the mean level spacing $\Delta = \langle \omega_{n+1} - \omega_n \rangle /(2\pi)$ given by Weyl's law $\Delta = c_0^3/(8\pi V f^2) \sim 0.54$~MHz. The very small value of $\Delta\Gamma$ evidences that our optimized system is extremely close to the CPA condition. In Fig.~\ref{fig:time-delay}e we observe a strong enhancement of the time delay for two representative realizations of CPA; $|\text{Re}[\tau(\omega)]|$ reaches values as high as 80~$\mu$s. For comparison, wavefronts that are orthogonal to $\psi_{\text{CPA}}$ yield an average time delay of 10.9~ns, almost four orders of magnitude smaller. The two CPA realizations in Fig.~\ref{fig:time-delay}d,e associated with a positive (negative) time delay correspond to a zero of $S$ just above (below) the real frequency axis. During the optimization of the meta-atom configurations, the zero can even cross the real frequency axis, an example thereof being shown in Fig.~\ref{fig:time-delay}f for which $\text{Re}[\tau(\omega)]$ suddenly jumps from $-25$~$\mu$s to 41~$\mu$s after a single iteration.

\begin{figure*}
    \centering
    \includegraphics[width=18cm]{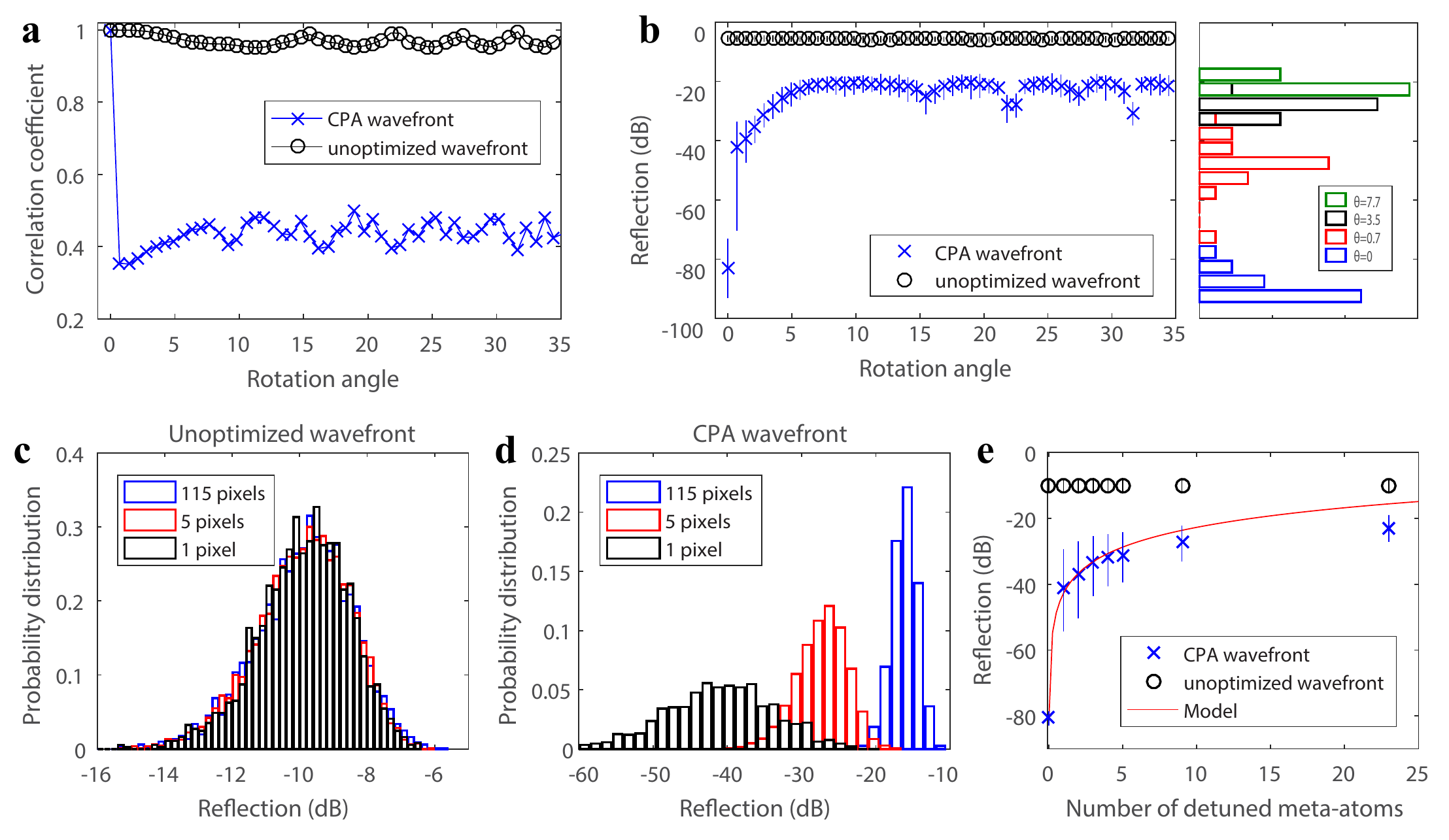}
    \caption{\textbf{Enhanced sensing with the time-delay singularity at the CPA condition. a,} Magnitude of the correlation coefficient $C(\psi_{in},\theta)$ of the outgoing field as a function of the size of the perturbation (angle of rotation $\theta$) for the case of injecting the CPA wavefront or an unoptimized wavefront. \textbf{b,} Corresponding multichannel reflection coefficient. The right panel shows a histogram of observed reflection values for different perturbation strengths, based on 18 CPA realizations with different initial orientations of the perturber.  \textbf{c, d, } Distributions of $R$ shown for a random incoming wavefront (\textbf{c}) and the CPA wavefront (\textbf{d}) for detuning 1, 5 and 115 meta-atoms.
    \textbf{e,} Variations of the reflection coefficient on a logarithmic scale at the CPA condition (blue crosses) and for a random incoming wavefront (black circles), as a function of the number of detuned meta-atoms $p$. The bars give the corresponding 90 percentiles for $R$. For small perturbations, the result obtained with the CPA wavefront is well explained by Eq.~(\ref{eq:R}) upon substituting $\Delta\omega = 2\pi K p$, with $K = 0.35$~MHz (see main text for details).}
    \label{fig:sensing}
\end{figure*}

\section{Time Delay Singularity for Optimal Sensitivity}

Now that we have have established and experimentally confirmed the physical origin of the time-delay singularity at the CPA condition in complex scattering media, we go on to investigate experimentally how this singularity can enhance the sensitivity of measurements to parametric perturbations, which is essential in sensing applications. The term ``sensitivity'' in the present work refers to a transduction coefficient of the sensor from the quantity to be measured (a perturbation $\Delta \alpha$ of a parameter $\alpha$) to an intermediate output quantity. This should not be confused with the smallest measurable change of the input quantity which pivotally depends on the measurement noise~\cite{langbein2018no}. In the following, first, we show that at the CPA condition, the ability to detect tiny perturbations is enhanced due to a rapid field decorrelation, the latter being intimately linked to the time delay. Then, second, we investigate the extent to which the CPA condition also enables a characterization of the perturbation in terms of its strength.

Wave chaos is generally known to be quite sensitive to perturbations; techniques known as diffuse wave spectroscopy (DWS) study the decorrelation of the outgoing field due to a perturbation in order to quantify the latter~\cite{pine1988diffusing,maret1997diffusing,de2003field}. Purely relying on the sensitivity of wave chaos corresponds (in our system tuned to have a real-valued zero) to injecting an unoptimized random wavefront $ \psi_{\text{in}} = \psi_{\text{rand}}$. However, because of the divergence of the delay time, the perturbation-induced decorrelation of the wave may be dramatically enhanced by injecting the optimized wavefront $ \psi_{\text{in}} = \psi_{\text{CPA}}$. We begin by defining a correlation coefficient based on the outgoing field as 

\begin{equation}
    C(\Delta \alpha) = \frac{\psi^\dagger_{out}(\alpha+\Delta \alpha) \psi_{out}(\alpha)}{\sqrt{R(\alpha) R(\alpha + \Delta \alpha)}}.
\end{equation}

\noindent To confirm the predicted rapid decorrelation of the outgoing field at the CPA condition experimentally, we gradually perturb the system by rotating a small metallic structure (a metallic pillar located on a metallic platform) in steps of $\Delta\theta = 0.62^{\circ}$ (see Fig.~\ref{fig:sensing}a). At each step, we measure $S(\theta)$ and evaluate the outgoing field $\psi_{\text{out}}(\theta) = S(\theta) \psi_{\text{in}}$ for unoptimized and optimized wavefronts. We repeat this procedure for 18 realizations of different initial positions of the rotating object. In Fig.~\ref{fig:sensing}a we plot $|C(\Delta\alpha)|$ as a function of the size of the perturbation (here the angle of rotation $\theta$). For $\psi_{\text{in}}=\psi_{\text{rand}}$, the field barely decorrelates even after 100 rotation steps. In other words, DWS-based sensing would not be capable of detecting or quantifying the perturbation that we consider. In contrast, for $\psi_{\text{in}}=\psi_{\text{CPA}}$, the field strongly decorrelates even for the smallest step: $C(\Delta \theta)\sim 0$. Fig.~\ref{fig:sensing}a therefore evidences the extreme sensitivity of the CPA condition for precision sensing in complex systems, well beyond that achievable with traditional DWS.

Having clarified that the rapid decorrelation of the outgoing field enhances the ability to detect the presence of a perturbation, we now explore to what extent this perturbation can also be characterized. For a chaotic cavity (not tuned to have a real-valued zero) and/or a random incoming wavefront, the average reflection coefficient is statistically independent of the perturbation strength. In contrast, at the CPA condition, $R(\alpha +\Delta \alpha)$ increases with $\Delta \alpha$ so that it may be possible to discriminate between two perturbations $\Delta \alpha_1$ and $\Delta \alpha_2$ based on measurements of $R(\Delta \alpha_1)$ and $R(\Delta \alpha_2)$. The reflection coefficient $R(\theta,\psi_{\text{CPA}})$ shown in Fig.~\ref{fig:sensing}b increases rapidly with $\theta$ but then saturates for $\theta > 5^{\circ}$. Note that the plateau reached for large angles is 20 dB below $R(\theta,\psi_{\text{rand}})$ as the perturbation is small. 

In wave-chaotic systems as our complex scattering enclosure, the energy density (and hence $\partial_\alpha \omega_n$ and $R$) are distributed quantities which fluctuate for different realizations of the system; an example are the histograms of $R(\theta,\psi_{\text{CPA}})$ shown in Fig.~\ref{fig:sensing}b. To investigate the dependence of $R(\alpha + \Delta\alpha)$ on the perturbation strength $\Delta\alpha$, a statistical analysis is hence required. We conveniently achieve this by considering a different type of perturbation: the detuning of $p$ meta-atoms away from the CPA configuration~\cite{del2020PA}. We successively determine $R$ for 200 realizations of $p$ detuned meta-atoms in 15 CPA realizations, with $p$ varying from $p=1$ to $p=115$. 

As expected, for random wavefronts, $R(p,\psi_{\text{rand}})$ is statistically independent of $p$: the distributions $P(R(p,\psi_{\text{rand}}))$ found for $p=1$, 5 and 115 detuned meta-atoms hence completely overlap, as seen in Fig.~\ref{fig:sensing}c. In contrast, Fig.~\ref{fig:sensing}d reveals that $P(R(p,\psi_{\text{CPA}}))$ strongly depends on $p$ because the variations of $\omega_n$ due to local changes of the boundary conditions increase with the number of detuned meta-atoms. We find that in the regime of small perturbations (here $p \leq 5$), $\langle R(p,\psi_{\text{CPA}}) \rangle$ is in good agreement with Eq.~(\ref{eq:R}) upon replacing the frequency shift $\Delta\omega = \omega - \omega_n$ with $2\pi K p$, where $K=0.35$~MHz is a constant depending on the scattering cross-section of the meta-atoms and the volume of the cavity. Our model's validity is confirmed by its faithful fit to the experimental data for $p\leq 5$ in Fig.~\ref{fig:sensing}e.

Nonetheless, we point out that $R$ being a distributed quantity results in fundamental limits on the precision with which the strength of any given perturbation can be determined. Indeed, the distributions $P(R(p,\psi_{\text{CPA}}))$ partially overlap for close values of $p$, as shown in Fig.~\ref{fig:sensing}d. This limitation may be understood as the price for the enhanced sensitivity to any perturbation within the system irrespective of its location.

\section{Discussion and Conclusion}

The discussed CPA condition can be interpreted as a special case of two distinct more general scattering phenomena: coherently enhanced absorption (CEA,~\cite{Chong2011}) and virtual perfect absorption (VPA,~\cite{baranov2017coherent}). 

CEA is a route to achieving very low (but finite) reflection values over an extended frequency range. Similarly to CPA, CEA relies on injecting the incoming wavefront giving the smallest reflection, which is the eigenstate of the matrix $S^\dagger (\omega) S(\omega)$ with minimal eigenvalue. Unless a zero of $S$ happens to lie on the real-frequency axis at $\omega$, this eigenstate generally does not correspond to an eigenstate of $S$ (CPA condition) so that the reflection coefficient does not vanish. Multiple resonances of the system are then involved and the reflection dip associated with CEA is not as pronounced as for CPA. In the case of CEA, the reflection coefficient decorrelates on a scale inversely proportional to the absorption mean free path \cite{Chong2011}. The time delay of waves and the sensitivity of the medium to a perturbation are hence generally bounded for CEA.

The idea of CPA is to bring a zero onto the real-frequency axis such that it can be accessed with a monochromatic excitation oscillating at a real frequency. If, however, the zero is not on the real frequency axis, it can still be accessed in the transient regime using a nonmonochromatic signal oscillating at a complex frequency. This VPA concept was recently studied for regular (almost) lossless systems for which the zero always lies in the upper half of the complex frequency plane~\cite{baranov2017coherent,trainiti2019coherent}. Consequently, the excitation signal has to exponentially increase in time to interfere destructively with the waves reflected off the system. The interaction of the incident pulse with the scattering medium then provides ideal energy storage until the interruption of the exponential growth of the injected signal. Given the generality of the scattering matrix formalism, this concept can be extended to disordered lossy matter such as complex scattering enclosures. The zeros may then lie anywhere in the complex frequency plane, implying that the necessary excitation is not always an exponentially increasing one. Interesting links with the sign of the time delay as discussed in the present work then arise.

To summarize, we have proposed a theoretical description of the hallmark sign of CPA in complex scattering system and we verified our theory experimentally. Our work rigorously explains the divergence of the time delay at the CPA condition and how this singularity justifies the optimal sensitivity of the CPA condition for detecting minute perturbations. This feature will enable novel precision sensing tools but also impact other areas such as filter applications and secure wireless communication~\cite{del2020PA,chen2020perfect}. Furthermore, our experiments demonstrated how a CPA condition can be accessed ``on demand'' at an arbitrary frequency without controlling the level of attenuation in the system. Finally, we note that our results are very general in nature and apply to other types of wave phenomena, too.

\textit{Note added.}~---~In the process of finalizing this manuscript, we became aware of related work~\cite{Frazier2020cpa} that also generalizes the concept of ``on-demand'' access to a real-valued scattering matrix zero in a complex scattering enclosure from single-channel~\cite{del2020PA} to multi-channel excitation.

\section*{Appendix: Experimental Methods}

\subsection{Experimental Setup}
Our complex scattering enclosure (depicted in Fig.~1a) is a metallic cuboid ($50~ \times 50~ \times 30~\text{cm}^3$) with two hemispheres on the inside walls to create wave chaos. In short, the difference between the trajectories of two rays launched from the same position in slightly different directions will increase exponentially in time. The system is excited via eight antennas (waveguide-to-coax transitions designed for operation in the $4<\nu<7$~GHz range) which are connected to an eight-port vector network analyser (VNA). The VNA acquires the full $8\times 8$ scattering matrix in one go. It emits signals at 0~dBm and operates with an intermediate-frequency bandwidth of 2~kHz to ensure a high signal-to-noise ratio. In the vicinity of the targeted CPA frequency $\nu_0 = 5.147$~GHz, we measure the spectra with an extremely small frequency step ($\Delta \nu=1$~kHz) and subsequently fit them with a linear regression in the Argand diagram to further reduce the impact of noise.

In order to tune our random system's scattering matrix, two programmable metasurfaces~\cite{cui2014coding} are placed on two neighboring walls of the cavity~\cite{dupre2015wave}. Each metasurface consists of an array of 152 1-bit programmable meta-atoms. Each meta-atom has two digitalized states, ``0'' and ``1'', with opposite electromagnetic responses. The working principle relies on the hybridization of two resonances of which one is tunable via the bias voltage of a pin diode; details can be found elsewhere~\cite{kaina2014hybridized}. The two states are designed to mimic Dirichlet and Neumann boundary conditions and their detailed characteristics are provided in the SM.

\subsection{Optimization of Meta-Atom Configurations}

Identifying a configuration of the programmable metasurfaces that yields CPA at the targeted frequency is a non-trivial task since there is no forward model describing the impact of the metasurface configuration on $S$. Hence, we opt for an iterative optimization algorithm similar to the one in Ref.~\cite{del2020optimal}. We begin by measuring $S$ for 200 random configurations. We then use the configuration for which the smallest eigenvalue $\lambda_8$ of $S(\nu_0)$ is the lowest as starting point. For each iteration, we randomly select $z$ meta-atoms and flip their state. If the resulting $S(\nu_0)$ has a lower $\lambda_8$, we keep the change. We gradually reduce the number of meta-atoms whose state is flipped per iteration, according to $z=\text{max}( \text{int} (50 e^{-0.02k}),1 )$, where $k$ is the iteration index.

\subsection{Characterization of Chaotic Cavity}

We characterize our system's linewidth associated with the $N$ attached channels, $N\gamma_c$, and its linewidth associated with global absorption effects, $\gamma_a$, in the following. To that end, we measure $S(\omega)$ for 300 random metasurface configurations and repeat the measurements for a number of channels connected to the cavity varying from $N=3$ to $N=8$. In each case, we estimate the average linewidth $\langle \gamma \rangle $ by fitting the exponential decay of average reflected intensities in the time domain, $I(t) = e^{-2\pi \langle \gamma \rangle t}$ (see SM for details). Using the variations of $\langle \gamma \rangle$ with respect to the number of attached channels,  $\langle \gamma \rangle = \langle \gamma_a \rangle + N \langle \gamma_c \rangle$, we obtain an average absorption strength $\langle \gamma_a \rangle =  11.2$~MHz and a linewidth associated with the channels $\langle \gamma_n \rangle = 8 \langle \gamma_c \rangle = 0.33$~MHz. 
Compared with the linewidth $\gamma_{\text{CPA}}$ found at the CPA condition it appears at first sight surprising that $\gamma_{\text{CPA}}$ exceeds $\langle \gamma_a \rangle$ and $\langle \gamma_n \rangle$ by more than one order of magnitude. 
The apparent paradox is resolved by noting that the resonance widths in wave-chaotic systems are not normally distributed; instead they have a distribution skewed toward lower values with a long tail for larger values~\cite{kottos2005statistics,kuhl2008resonance}. Such a distribution is not fully characterized by its average, and observing values well above the average, as in the case of $\gamma_{\text{CPA}}$, is by no means impossible.

\section*{Acknowledgments}
\noindent This publication was supported by the French ``Agence Nationale de la Recherche'' under reference ANR-17-ASTR-0017, by the European Union through the European Regional Development Fund (ERDF), and by the French region of Brittany and Rennes M{\'e}tropole through the CPER Project SOPHIE/STIC \& Ondes. The metasurface prototypes were purchased from Greenerwave. The authors acknowledge P.~E.~Davy for the 3D rendering of the experimental setup in Fig.~1a. M. D. acknowledges the Institut Universitaire de France.


\bibliographystyle{apsrev4-1}


%

\clearpage

\section*{SUPPLEMENTAL MATERIAL}
\renewcommand{\thefigure}{S\arabic{figure}}
\renewcommand{\theequation}{S\arabic{equation}}
\setcounter{equation}{0}
\setcounter{figure}{0}
\setcounter{section}{0}

\section{Theoretical Model}

In this section, we provide additional details and intermediate algebraic steps for our theoretical results reported in the main text.

\subsection{Effective Hamiltonian approach}
We analyze the properties of the scattering matrix $S(\omega)$ using a non-perturbative approach based on the effective Hamiltonian. This model was initially developed to characterize open quantum systems (see Ref.~\cite{Okolowicz2003}) but also qualitatively describes the transport of classical waves in chaotic systems \cite{Kuhl2005}. The coupling of the channels to a system is analyzed in terms of an effective Hamiltonian $H_{\text{eff}}$ expressed as $H_{\text{eff}} = H_0 - iVV^T/2$. The Hermitian Hamiltonian $H_0$ characterizes the closed system and $V$ is a real matrix of dimensions $M \times N$ describing the coupling of the $N$ channels to the $M$ modes of the closed system. 
To compare experimental results with quantum-mechanical predictions, we use the analogy between the experimental eigenfrequencies $\omega_n$ and linewidths $\Gamma_n$ with their quantum-mechanical counterparts $E_n = \omega_n^2$ and $\Tilde{\Gamma}_n = \omega_n \Gamma_n$ \cite{Kuhl2005}. Under the assumption of small linewidths, $\Gamma_n / \omega \ll 1$, we can express the scattering matrix giving the field coefficient between incoming and outgoing channels as

\begin{equation}
    S(\omega) = \id - iV^T \frac{1}{\omega\id - H_0 + \frac{i}{2}(VV^T + \Gamma_a \id)} V .
    \label{eq:S}
\end{equation}

\noindent Here, we have introduced the linewidth $\Gamma_a$ to take into account the losses within  the internal system. One can alternatively express the scattering matrix in terms of the eigenfunctions of the non-effective Hamiltonian. In absence of absorption, the poles of $S$ are found at the complex eigenvalues $\Tilde{\omega}_m = \omega_m - i\Gamma_m/2 $. Here $\omega_m$ is the central frequency and $\Gamma_m$ is the associated linewidth. The scattering matrix $S(\omega)$ can be decomposed as the superposition of natural resonances: 

\begin{equation}
    S(\omega) = \id - i \Sigma_{m=1}^M \frac{W_mW_m^T}{\omega-\omega_m +i(\Gamma_m + \Gamma_a)/2} .
\label{eq:S_eigenfunctions}
\end{equation}

\noindent The vectors $W_m$ are the projection of the eigenfunctions $\phi_n$ onto the channels coupled to the system: $W_m = V^T\phi_m$.

The zeroes $z_m$ of $S(\omega)$ are the complex frequencies satisfying $\text{det}S(z_m) = 0$. They are found using the relation~\cite{fyodorov2017distributionB}

\begin{equation}
    \text{det} S(\omega) = \frac{\text{det}(\omega \id - H_0 +i\Gamma_a - i\Gamma)}{\text{det}(\omega \id - H_0 +i\Gamma_a + i\Gamma)},
    \label{eq:det}
\end{equation} 

\noindent where $\Gamma$ is the matrix $\Gamma = V^T V/2$. Eq.~(\ref{eq:det}) shows that in absence of absorption ($\Gamma_a = 0$) the zeroes and poles of the scattering matrix are symmetrically placed in the upper and lower complex plane with $z_m = \Tilde{\omega}_m^*$. When absorption within the medium is added ($\Gamma_a \neq 0$), the zeroes and poles move in the complex plane with $z_m = \omega_m +i(\Gamma_m - \Gamma_a)/2$ and $ \Tilde{\omega}_m = \omega_m -i(\Gamma_m + \Gamma_a)/2$.

CPA can be realized when one zero of $S(\omega)$ is located on the real frequency axis. For notational ease, we label the $m$th resonance (out of the $M$ resonances contributing to the sum in Eq.~(\ref{eq:S_eigenfunctions})) that is associated with this special real-valued zero with the subscript $n$. This notation is applied to all quantities associated with individual resonances, such as $\omega_m$ or $\Gamma_m$. 
Eq.~(\ref{eq:det}) demonstrates that CPA is hence found when  
the dissipation within the cavity fully compensates losses through channels for a given pole: $\Gamma_n = \Gamma_a$. The frequency of this CPA condition is then $\omega = \omega_n$. 

\subsection{Perfectly absorbed incoming wavefront}

When a zero of the scattering matrix crosses the real frequency axis, there exists a vector $\psi_{CPA}$ which satisfies $S(\omega_n)\psi_{CPA} = 0$. In analogy with random lasing, this vector has been identified as the time-reverse of the lasing wavefront. We now demonstrate that $\psi_{CPA}$ indeed is the phase-conjugate of the modal vector, $\psi_{CPA} = W_n^* / \| W_n \|$. Using Eq.~(\ref{eq:S_eigenfunctions}), we find that the outgoing vector $\psi_{out} = S\psi_{CPA}$ for $\omega = \omega_n$ and $\Gamma_a = \Gamma_n$ is

\begin{equation}
    \psi_{out} = \frac{1}{\| W_n \|}\left[W_n^*- i\Sigma_{m=1}^M W_m \frac{W_m^T W_n^*}{\Tilde{\omega}_n^* - \Tilde{\omega}_m}\right] .
\label{eq:psi_out1}
\end{equation}

\noindent The eigenfunctions $\phi_m$, and hence the vectors $W_m$, are generally non-orthogonal in open systems: $\phi_{m_1} ^\dagger \phi_{m_2} \neq \delta_{m_1 m_2}$. They are, however, bi-orthogonal and the degree of correlation of the vectors $W_m$ is related to the correlation between eigenfunctions $ \phi_{m_1}^T \phi_{m_2}^*$ as

\begin{equation}
    \phi_{m_1}^T \phi_{m_2}^* = i\frac{W_{m_1}^T W_{m_2}^*}{\Tilde{\omega}_{m_1}^* - \Tilde{\omega}_{m_2}}.
\label{eq:correlation}
\end{equation}

\noindent Note that $\phi_{m_1}^T \phi_{m_2}^*$ is an element of the Bell-Steinberger non-orthogonality matrix $U$ giving the correlation between each pair of eigenfunctions of the system. By incorporating Eq. (\ref{eq:correlation}) within Eq.~(\ref{eq:psi_out1}) we get

\begin{equation}
    \psi_{out} = \frac{1}{\| W_n \|} [W_n^* - V^T (\Sigma_{m=1}^M \phi_m \phi_m^T) \phi_n^*].
\end{equation}

\noindent Using the completeness of the eigenfunctions, $\Sigma_{m=1}^M \phi_m \phi_m^T = \id$, and the relation $V^T \phi_n^* = W_n^* $, we finally verify that $\psi_{out} = 0$, as expected.

\subsection{Reflection coefficient}

\begin{figure}
    \centering
    \includegraphics[width =  \columnwidth]{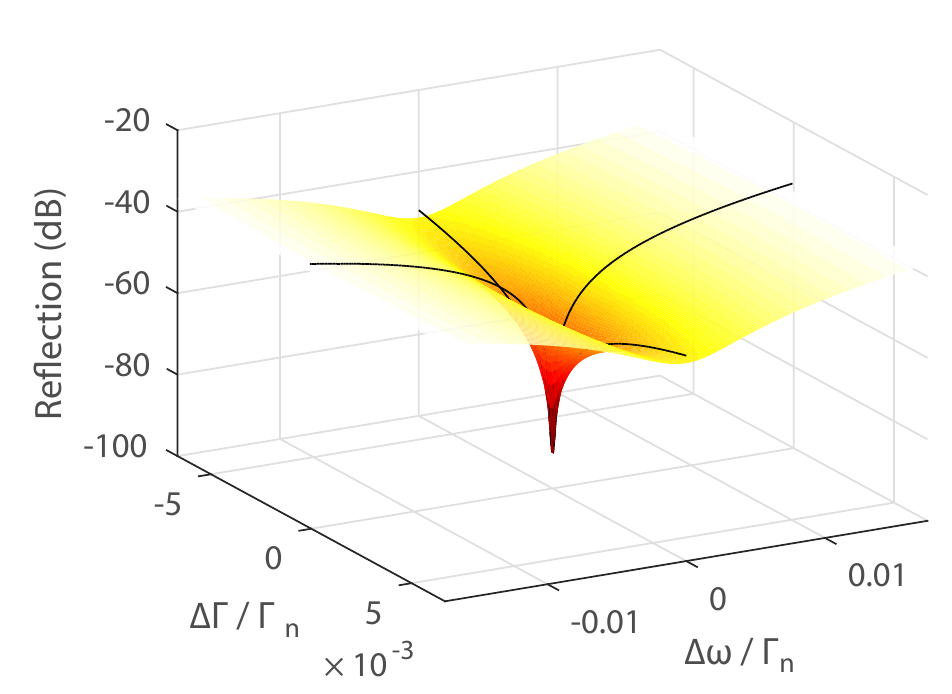}
    \caption{Simulation results of the reflection coefficient $R(\omega)$ in the vicinity of a CPA condition at $\omega_n$ are presented as a function the linewidth detuning $\Delta \Gamma_a / \Gamma_n$ and the frequency detuning $\Delta \omega / \Gamma_n$. The results in excellent agreement with the theoretical predictions (black lines) shown for $\Delta\omega = 0$ and $\Delta\Gamma = 0$.}
    \label{fig:RMT_model}
\end{figure}

In a realistic experiment we never perfectly achieve the CPA condition with $\omega = \omega_n$ and $\Gamma_a = \Gamma_n$. Therefore, we now set $\omega =\omega_n + \Delta\omega$ and $\Gamma_a = \Gamma_n +\Delta\Gamma$. We use a perturbation approach to derive the multichannel reflection coefficient $R(\omega) = \|S(\omega) \psi_{CPA} \|^2$. Eq. (\ref{eq:psi_out1}) yields

\begin{equation}
    \psi_{out} = \frac{1}{\| W_n \|}\left[W_n^*- i\Sigma_m W_m \frac{W_m^T W_n^*}{\Tilde{\omega}_n^* - \Tilde{\omega}_m +\Delta\omega +i\Delta\Gamma/2}\right] .
\label{eq:psi_out2}
\end{equation}

\noindent In the vicinity of the CPA condition ($\Delta\omega \ll (\omega_{n+1} - \omega_n)$ and $\Delta\Gamma \ll \Gamma_n$) we assume that the perturbation only affects the $n$th resonance. Eq.~(\ref{eq:psi_out2}) can then be simplified to

\begin{equation}
    \psi_{out} \sim \frac{W_n^*}{\| W_n \|}\left[1 - i\frac{\Gamma_n}{\Delta\omega+i( \Gamma_n + \Delta\Gamma/2)}\right] .
\label{eq:psi_out3}
\end{equation}

\noindent Straightforward calculations finally lead to

\begin{equation}
    R(\omega) = \|\psi_{out} \|^2 = \frac{4 \Delta \omega ^2 + (\Delta\Gamma)^2}{4 \Delta \omega ^2 + (2\Gamma_n+\Delta\Gamma)^2}.
\label{eq:R}    
\end{equation}

Finally, the noise level in the experiment corrupts the measurement of $S$ as $S\rightarrow S+\Delta S$, where $\Delta S$  is a random matrix with complex random elements with Gaussian distribution. This mismatch increases the reflection coefficient by an additional level of $\| \Delta S \|_F^2 / N$, where $\| \dots \|_F$ denots the Frobenius norm and $N$ is the number of channels coupled to the system. We hence obtain Eq.~(2) of the main text.

We now compare Eq.~(\ref{eq:R}) to numerical simulations of our model using Eq.~(\ref{eq:S}). $H_0$ is modelled as a real symmetric matrix of dimensions $500 \times 500$ drawn from the Gaussian orthogonal ensemble; the coupling matrix $V$ to the $N=8$ channels is modelled as a real random matrix with Gaussian distribution. The channels are assumed to be fully coupled to the system. We select the resonance which is nearest to the middle of the band and then explore the reflection coefficient $R = \| S(\omega) \psi_{CPA} \|^2  $ as a function of frequency detuning $\omega =\omega_n + \Delta \omega$ and linewidth detuning $\Gamma_a = \Gamma_n + \Delta \Gamma$. The results shown in Fig.~\ref{fig:RMT_model} demonstrate an excellent agreement between the random matrix simulations and our theoretical predictions.

\subsection{Time delay}
\begin{figure}
    \centering
    \includegraphics[width= \columnwidth]{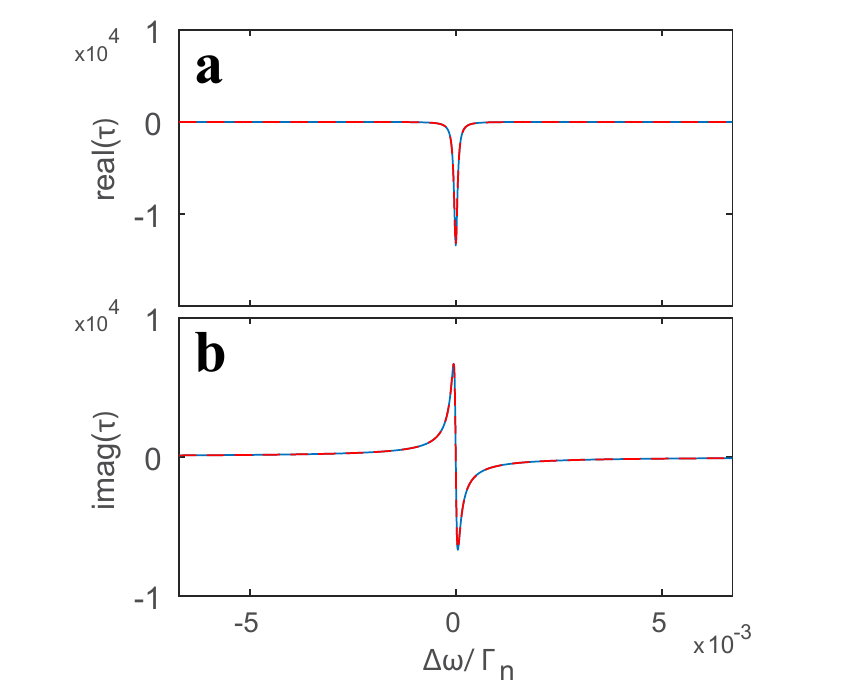}
    \caption{Simulation results (blue lines) of the real (a) and imaginary (b) parts of the time delay in the vicinity of a CPA condition at $\omega_n$ as a function of the frequency detuning $\Delta \omega / \Gamma_n$, in excellent agreement with the theoretical predictions (red dashed lines).}
    \label{fig:time_delay_rmt}
\end{figure}

The time delay can be expressed in terms of the Wigner-Smith (WS) operator $Q(\omega) = -i S(\omega)^\dagger \partial_\omega S(\omega)$:  

\begin{equation}
    \tau(\omega) = \frac{\psi_{CPA}^\dagger Q(\omega) \psi_{CPA}}{\psi_{CPA}^\dagger S(\omega)^\dagger S(\omega) \psi_{CPA}}. \label{eq:tau_WS}    
\end{equation}
 
\noindent Note that the projection of the incoming wavefront onto the Wigner-Smith operator is normalized by the reflection coefficient so that $\tau(\omega)$ can be interpreted as phase derivative.

Obviously a singularity may appear at the CPA condition with $\omega = \omega_n$ and $\Gamma_a = \Gamma_n$ as $R(\omega_n) = \psi_{CPA}^\dagger S(\omega_n)^\dagger S(\omega_n) \psi_{CPA} = 0$. However, this singularity is removed for $\omega \neq \omega_n$ and/or $\Gamma_a \neq \Gamma_n$. We now estimate $\tau(\omega)$ in the vicinity of the CPA using the same approach as for $R(\omega)$. We first calculate the derivative of $S(\omega)$ with respect to $\omega$:

\begin{equation}
    \partial_\omega S(\omega) = i\Sigma_{m=1}^M \frac{W_m W_m^T}{[\omega-\omega_m +i(\Gamma_m/2 + \Gamma_a/2)]^2}.
\end{equation}

\noindent This leads to

\begin{equation}
    \psi ^\dagger Q(\omega) \psi = \frac{4\Gamma_n}{4\Delta\omega^2 + (2\Gamma_n + \Delta\Gamma)^2} \frac{(2\Delta \omega -i \Delta \Gamma)}{2\Delta \omega +i (2\Gamma_n + \Delta \Gamma)}
\end{equation}

\noindent As the denominator of Eq.~(\ref{eq:tau_WS}) is given by Eq.~(\ref{eq:R}), straightforward calculations finally yield the real and imaginary parts of $\tau$:

\begin{equation}
    \text{Re}[\tau] = \frac{4\Gamma_n}{4\Delta\omega^2 + \Delta\Gamma^2} \frac{4\Delta\omega^2 - \Delta\Gamma (2\Gamma_n + \Delta\Gamma)}{4\Delta\omega^2 + (2\Gamma_n + \Delta\Gamma)^2}
\label{eq:tau_r}    
\end{equation}

\noindent and

\begin{equation}
    \text{Im}[\tau] = -\frac{4\Gamma_n}{4\Delta\omega^2 + \Delta\Gamma^2} \frac{4\Delta\omega (\Gamma_n + \Delta\Gamma) }{4\Delta\omega^2 + (2\Gamma_n + \Delta\Gamma)^2}
\label{eq:tau_i}    
\end{equation}

\noindent Interestingly, at the resonance with the mode ($\Delta\omega = 0$) a phase transition appears on the real part of $\tau$ which is positive for $\Gamma_n > \Gamma_a$, diverges for $\Gamma_n = \Gamma_a$  and negative for $\Gamma_n < \Gamma_a$. This theoretical prediction is fully confirmed by the agreement with our model in Fig.~(2) of the main text and Fig.~\ref{fig:time_delay_rmt}.

\subsection{Negative delay time}

\begin{figure}
    \centering
    \includegraphics[width =  \columnwidth]{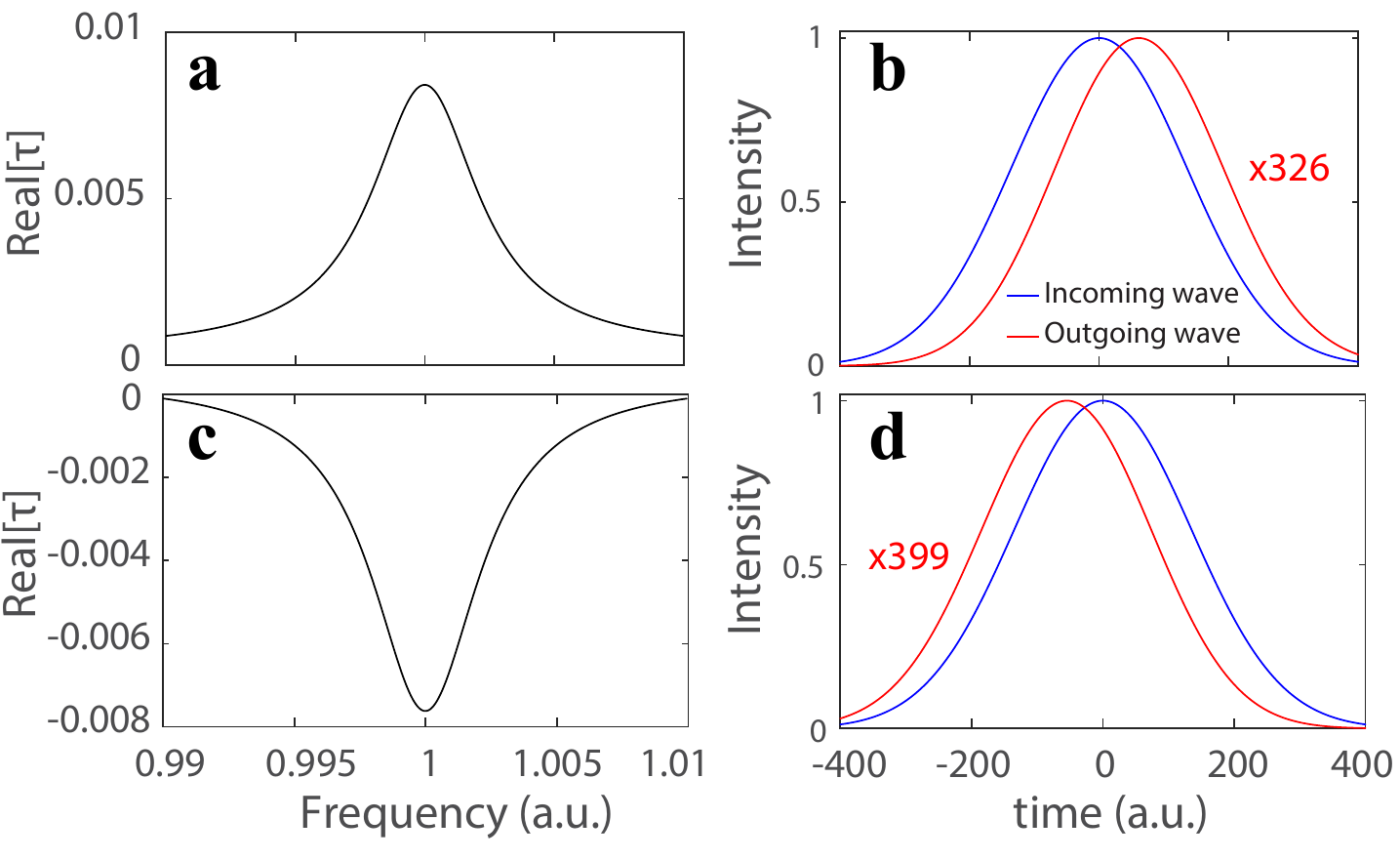}
    \caption{Simulation result of the time delay (a,c) and the incoming and outgoing intensity in the time domain (b,d) in the vicinity of a CPA for (a,b) a positive time delay and (c,d) a negative time delay. The temporal intensities are found from an inverse Fourier transform of the outgoing field $\psi_{out}(\omega)$ for an incoming Gaussian pulse. The width of this incident pulse in the frequency domain is 0.2\% of the central frequency. As the incoming wave is strongly absorbed, the outgoing intensities are normalized by factors of 326 in b and 399 in d.}
    \label{fig:1D}
\end{figure} 

One suprizing result is that the time-delay can be negative. We illustrate this effect by computing the outgoing intensity in the time domain for an incident pulse with a very small bandwidth. For the sake of simplicity, we realize a single channel-simulation ($N=1$) using the effective Hamiltonian approach. We obtain the spectrum of the scattering matrix in the vicinity of a CPA condition for $\Gamma_a < \Gamma_n$ and then for $\Gamma_a > \Gamma_n$. In the first case, the delay time is positive (see Fig.~\ref{fig:1D}a) while it is negative in the second case (see Fig.~\ref{fig:1D}c). We obtain the outgoing intensity in the time domain using an inverse Fourier transform of the reflection coefficient. As expected, we observe that the peak of the pulse is shifted positively (negatively) for positive (negative) time delays (see Fig.~\ref{fig:1D}b,d). We highlight, however, that the outgoing intensity has been normalized to visualize this effect. As the incoming wave is strongly absorbed within the system, the normalization factors of the outgoing wave relative to the incoming wave are 326 and 399 in these two cases.

Obviously, this effect does by no means break causality. In the case of strongly absorbed signals, negative time delays for $\Gamma_a > \Gamma_n$ arise due to the distortion of the incident pulse in resonant absorbing systems~\cite{muga1998negativeB,muga2002bounds,tanaka2003propagationB,van2008nonthreshold,Durand2019B}. The signal at long times is indeed more strongly absorbed than the signal at early times. As a consequence, the maximum of the outgoing pulse is shifted to shorter times and the pulse appears to be delayed negatively. For $\Gamma_a < \Gamma_n$, the effect is inverted so that the peak is delayed to longer times. This effect hence relies on temporal variations of the absorption of the signal.

\subsection{Connection with the \textit{generalized} Wigner-Smith operator}

In this section, we prove that the incoming wavefront $\psi_{\text{CPA}}$ is also the eigenstate with maximal eigenvalue of the \textit{generalized} Wigner-Smith operator $Q'_\alpha$ for a perturbation $\alpha$ defined  as $Q'_\alpha = -iS^{-1} \partial_\alpha S $~\cite{Ambichl2017bB}. Note that here that we distinguish between $Q'_\alpha $ and $Q_\alpha = -iS^{\dagger} \partial_\alpha S $. These two operators coincide for a unitary scattering matrix which is not the case here due to absorption. 

First, we estimate the pseudo-inverse of the scattering matrix using a singular value decomposition of $S$: $S = \Sigma_{n=1}^N u_n \lambda_n v_n^\dagger$. This leads to 
\begin{equation}
    S^{-1} = \Sigma_{n=1}^N v_n \left(\frac{1}{\lambda_n}\right) u_n^\dagger.
\label{eq:inverse_of_S}    
\end{equation}

\noindent $u_n$ and $v_n$ are the left and right singular vectors of $S$ associated the singular values $\lambda_n$. In the vicinity of a CPA condition, one singular value of the the scattering matrix is vanishing, $R(\omega) = \lambda_N^2 \rightarrow 0$, for an incoming wavefront $v_n = \psi_{\text{CPA}} $. The singular vector $u_n$ is the normalized outgoing field $\psi_{out}/\|\psi_{out}\|$  given by Eq.~(\ref{eq:psi_out3}) with a global phase shift corresponding to the phase of the term $\left[1 - i\frac{\Gamma_n}{\Delta\omega+i( \Gamma_n + \Delta\Gamma/2)}\right]$.

The pseudo inverse $S^{-1}$ in the vicinity of the CPA condition can then be approximated by $S^{-1} = v_N (1/\lambda_n) u_n^\dagger$. Using that $S \psi_{\text{CPA}} =  \psi_{\text{out}}$ with $\| \psi_{\text{out}} \|^2 = R$, the pseudo-inverse can also be expressed as 

\begin{equation}
    S^{-1}(\alpha) = \frac{\psi_{\text{CPA}} \psi_{\text{out}}^\dagger}{R(\alpha)} .
\end{equation}

\noindent As $S^{-1}$ is a matrix of unit rank, the \textit{generalized } Wigner-Smith operator $Q_\alpha = -iS^{-1} \partial_\alpha S $ is hence also of unit rank with a left eigenvector which is $v_n = \psi_{\text{CPA}} $. We have shown in Fig.~3a of the main text that the field-field correlation is extremely sensitive to the rotation of the perturber at the CPA condition. It comes therefore as no surprise that the wavefront yielding the optimal sensitivity to the perturbation is also the CPA wavefront.

\subsection{Simulations of a perturbation}

\begin{figure}
    \centering
    \includegraphics[width =  \columnwidth]{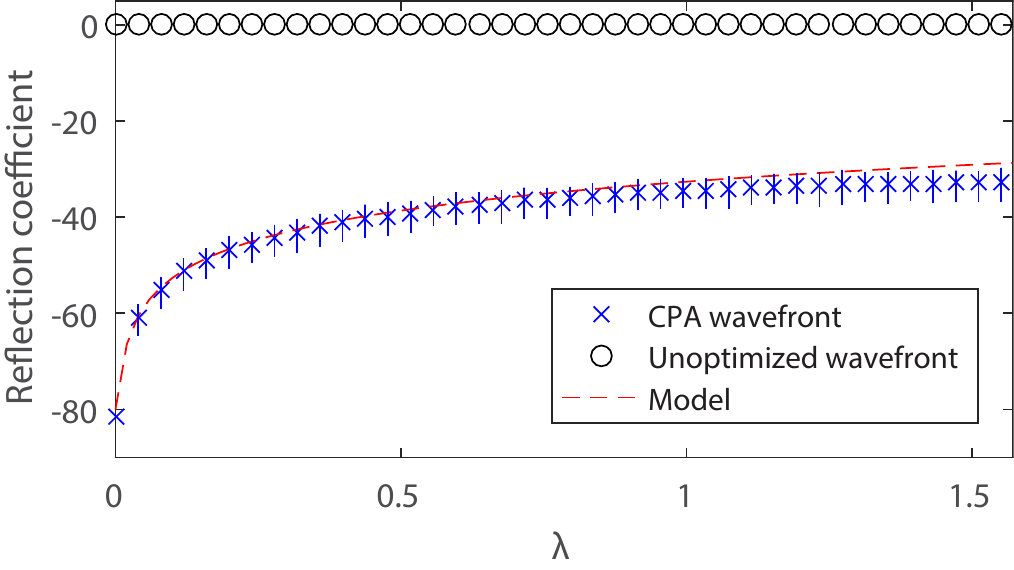}
    \caption{Variations of the reflection coefficient on a logarithmic scale at the CPA condition (blue crosses) and for a random incoming wavefront (black circles), as a function of the perturbation $\lambda$ in the effective Hamitonian model. The red line gives the best fit of $R$ with Eq.~(\ref{eq:R}) in which $\Delta\omega$ is replaced by $2\pi K \lambda$, where $K$ is a constant.}
    \label{fig:scaling_perturb_Heff}
\end{figure} 

In the main text, we find that the increase of the reflection coefficient at the CPA condition for a perturbation of $p$ meta-atoms is well predicted by Eq.~(1) of the main text in which $\Delta\omega$ is replaced by $2 \pi K p$ ($K$ is a constant). Here, we verify this scaling in random-matrix simulations based on the effective Hamiltonian. The perturbation is modelled by a variation of the initial Hamiltonian $H_0$ as 

\begin{equation}
    H = \text{cos}(\lambda) H_0 + \text{sin}(\lambda) \Delta H
\end{equation}

\noindent In this model, $\Delta H$ is a random Hamiltonian which is statistically independent of $H_0$. The parameter $\lambda$ determines the perturbation strength and the average density of states is preserved~\cite{haake_chapter,PhysRevE102010201}. The scattering matrix is computed from Eq.~(\ref{eq:S}). The reflection coefficient $R$ is finally obtained in simulations for the CPA wavefront $\psi_{\text{in}} = \psi_{\text{CPA}}$ and for a random incoming wavefront, $\psi_{\text{in}} = \psi_{\text{rand}}$ and $R$ is averaged over 200 CPA realizations in each case. The linewidth detuning $\Delta\Gamma$ is chosen so that $10\text{log10}(R) = -80$~dB for $\lambda = 0$.

We observe in Fig.~(\ref{fig:scaling_perturb_Heff}) a very good agreement between the simulation results and Eq.~(\ref{eq:R}) in which $\Delta\omega$ is replaced by $2\pi K \lambda$,

\begin{equation}
   R(\omega) = \frac{4 (2\pi K \lambda) ^2 + (\Delta\Gamma)^2}{4 (2\pi K \lambda) ^2 + (2\Gamma_n+\Delta\Gamma)^2} 
\end{equation}

\noindent Here $K$ is a constant which is adjusted to provide the best fit of $R$. Eq.~(\ref{eq:R}) is however valid only for small shifts of the central frequency $\omega_n$. A deviation between simulation results and the model is hence found for $\lambda > 0.5$.

\section{Experimental Details}

In this section, we provide further experimental details on our ``on-demand'' realization of CPA in a programmable complex scattering enclosure. We begin by providing additional details regarding working principle and key characteristics of our programmable meta-atoms. Then, we provide supplementary descriptions of our experimental setup, we characterize our chaotic cavity, and detail our data processing. Finally, we provide an additional analysis of the optimized metasurface configurations.

\subsection{Programmable Metasurfaces}

A programmable metasurface is an ultra-thin array of meta-atoms with programmable electromagnetic response. Programmable metasurfaces are emerging as powerful young member of the metamaterial family owing to their ability to dynamically shape electromagnetic fields as well as to the ease of fabricating the programmable metasurfaces. Initially, programmable metasurfaces were conceived for free-space control of waves~\cite{cui2014codingB,li2017electromagnetic} but more recently their powerful usefulness to tune the scattering properties of complex media has been discovered~\cite{dupre2015waveB}. This has led to a wide range of applications in indoor wireless communication~\cite{del2019optimally,di2019smart}, sensing~\cite{del2018precise} and even analog computing~\cite{del2018leveraging}.

In our experiments, we use a 1-bit programmable meta-atom as unit cell of which the programmable metasurfaces are composed. Each meta-atom has two digitalized states, ``0'' and ``1'', and can be configured individually to be in either of these states. The fundamental working principle is that introduced in Ref.~\cite{kaina2014hybridizedB}. The meta-atom consists of two resonators that hybridize, and the resonance frequency of one of the two can be altered by controlling the bias voltage of a PIN diode. This results in a phase change of roughly $\pi$ for the reflected wave.

\begin{figure}
    \centering
    \includegraphics[width =  \columnwidth]{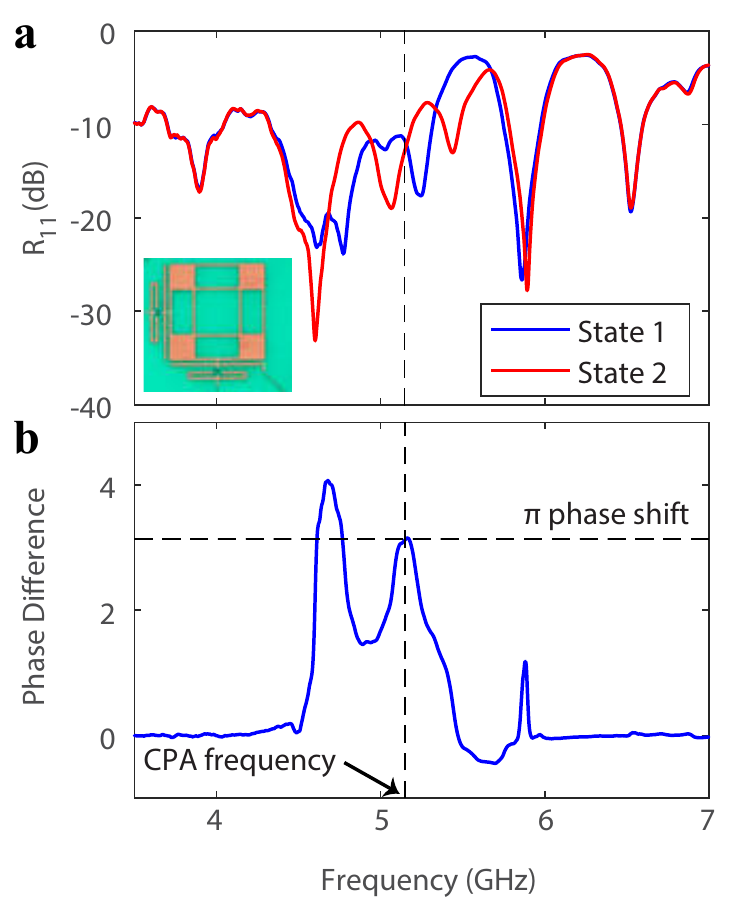}
    \caption{(a) Reflection coefficient $R_{11} = |S_{11}|^2$ measured with a horn antenna placed in front of the metasurface, for the two distinct states ``0'' and ``1''. (b) Phase difference of the $S_{11}$ parameter between the two states. The frequency chosen to obtain CPA by programming the metasurfaces correspond to that at which the metasurface responses can create a phase shift of $\pi$.}
    \label{fig:metasurface}
\end{figure}

Unlike the meta-atoms in Ref.~\cite{kaina2014hybridizedB}, our prototype offers independent control over the two orthogonal field polarizations. In short, this is achieved by fusing two single-polarization meta-atoms, one rotated by $90^{\circ}$ with respect to the other, into a single one. As seen in the inset of Fig.~\ref{fig:metasurface}a, the two meta-atoms share the same fixed resonator but one PIN diode controls the tunable resonator for each of the two polarizations.

Thoroughly characterizing the exact response of the meta-atoms is challenging since the characterization procedure inevitably impacts the results. Here, we illuminate the metasurface with a plane wave polarized along one of the two polarizations with a horn antenna. We synchronize the states of all meta-atoms for this measurement and measure the return loss of the horn antenna in close proximity to the metasurface for the two possible states, ``0'' and ``1''. The resulting curves are displayed in Fig.~\ref{fig:metasurface}. Recall that the absolute values of the return loss are modulated by the horn antenna's transfer function. The absolute value $R_{11} = |S_{11}|^2$ is therefore not an indicator of the amount of energy absorbed by the metasurface. 
The phase difference of $S_{11}$ for the two states, however, confirms that around the working frequency the metasurface does indeed offer a phase shift of roughly $\pi$. In other words, every meta-atom can be configured to mimic a Dirichlet or a Neumann boundary condition. Similar results are obtained for the other polarization.

\subsection{Experimental Setup}

\begin{figure}
    \centering
    \includegraphics[width =  \columnwidth]{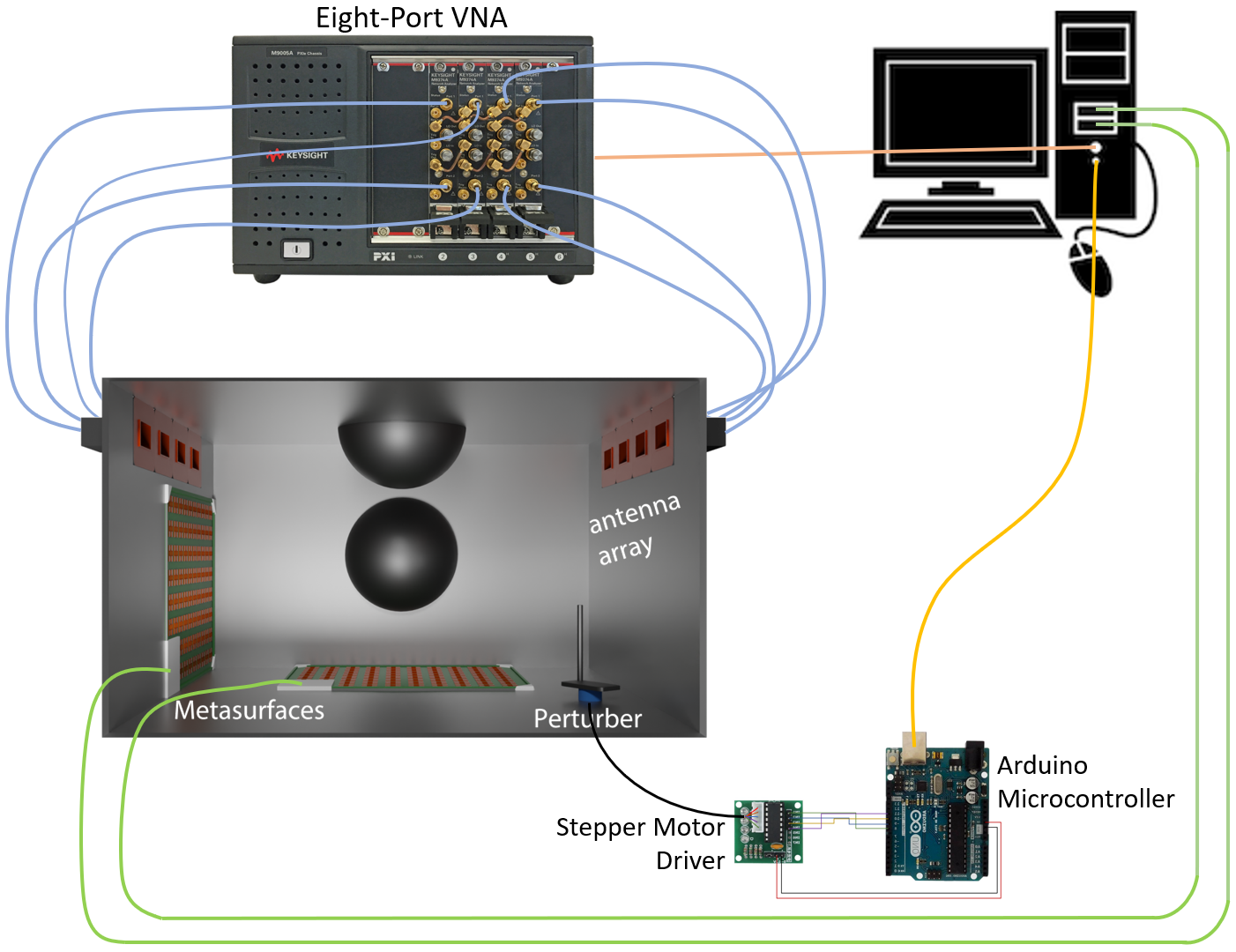}
    \caption{The schematic drawing illustrates how the complex scattering enclosure's eight ports are connected to a vector network analyzer (VNA) as well as how a computer programs the metasurfaces, controls the VNA and commands the stepper motor.}
    \label{fig:schematic_drawing}
\end{figure}

\begin{figure}
    \centering
    \includegraphics[width =  \columnwidth]{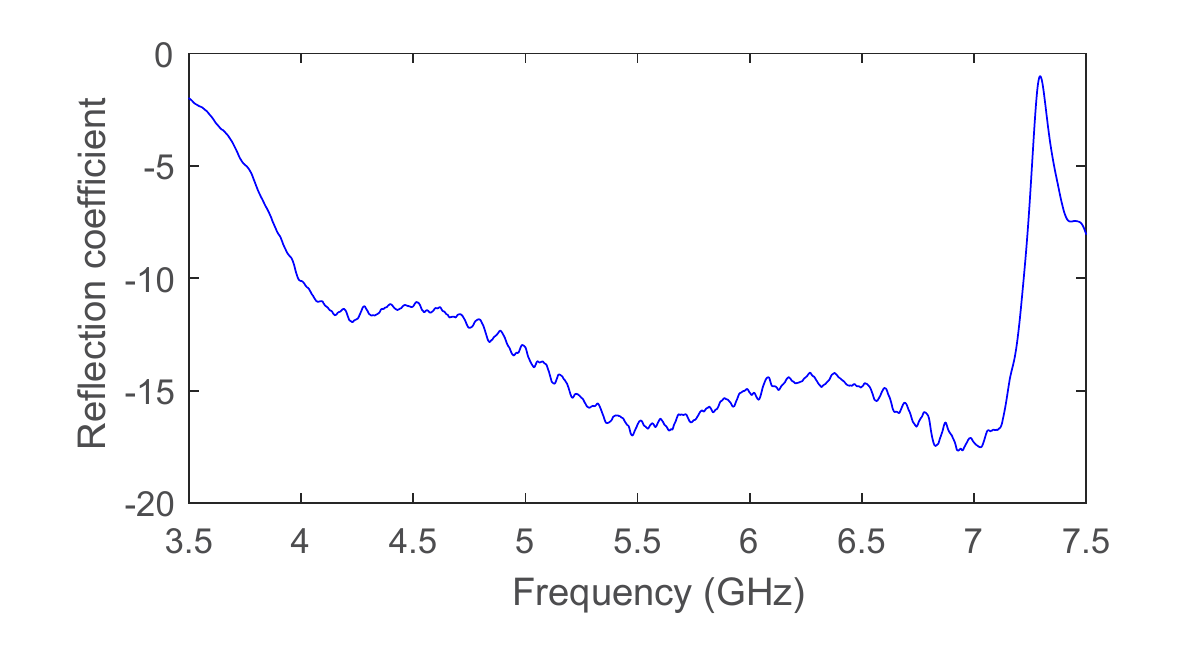}
    \caption{Reflection coefficient $|S_{11}|^2$~[dB] measured for a port (coax-to-waveguide transition) in free space.}
    \label{fig:antenna_FP}
\end{figure}

A global overview of our experimental setup is provided in Fig.~\ref{fig:schematic_drawing}. The complex scattering enclosure is a metallic cuboid of dimensions $50~ \times 50~ \times 30~\text{cm}^3$ with two hemisphere deformities on two mutually perpendicular walls. Two arrays of programmable meta-atoms with the characteristics detailed in the previous section are placed on two mutually perpendicular walls. The system is connected to the outside via eight channels. Specifically, the eight ports of our vector network analyzer (VNA, Keysight PXI) are connected via coaxial cables to eight identical waveguide-to-coax transitions. The latter are located in pairs of four on two sides of the cavity, but in a chaotic system the spatial location of the scattering channels is irrelevant. These ports are well adapted in the range between 4~GHz and 7~GHz, as evidenced by a free-space measurement of one port's reflection coefficient in Fig.~\ref{fig:antenna_FP}. A computer is connected to the VNA to initiate measurements of the $8\times 8$ scattering matrix. Both programmable metasurfaces are also connected to and configured by the computer. The stepper motor (28BYJ-48) is controlled by the computer via an Arduino microcontroller and a driver board (STP01 Joy-It). A metallic rod attached to a rotating metallic plate is fixed on the rotating motor to perturb the scattering system in small steps.

\subsection{High-Precision Measurements}

\begin{figure}[h]
    \centering
    \includegraphics[width =  \columnwidth]{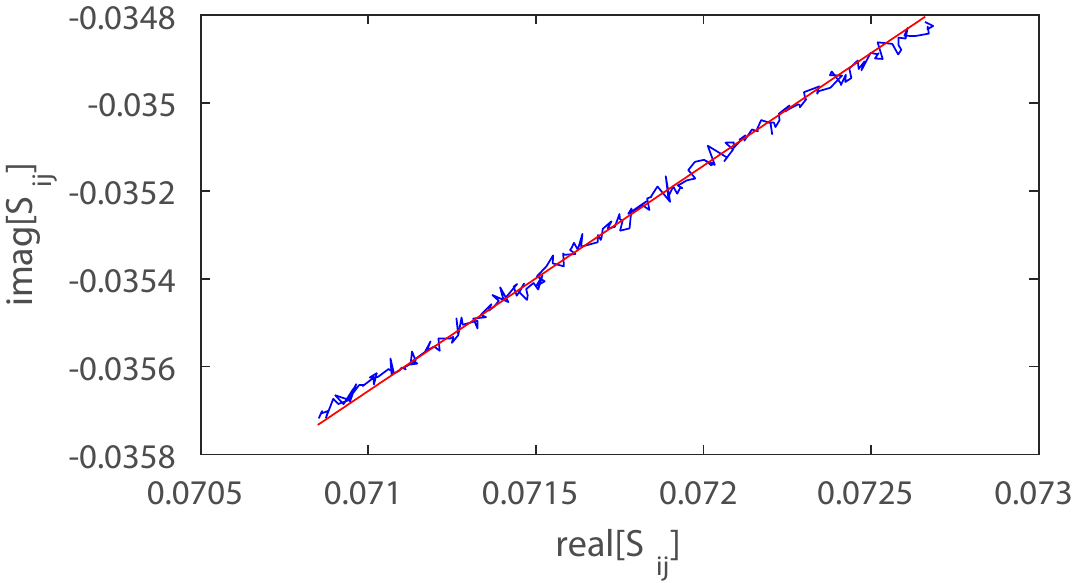}
    \caption{Example of a scattering coefficient in the vicinity of $\nu_0$ fitted in the complex Argand diagram by a linear function.}
    \label{fig:fit}
\end{figure}

Given the extreme sensitivity of the CPA condition to any sort of detuning, we take a number of measures to enhance the precision of our measurements. Obvious measures include a careful choice of settings of the vector network analyzer: an intermediate-frequency bandwidth of 1~kHz and an emitted power of 5~dBm. A more subtle additional technique that we employ is to measure $S(\nu)$ with an extremely small frequency step of 1~kHz in the vicinity of the targeted CPA frequency $\nu_0$. We then fit this data with a linear function in the complex Argand diagram~\cite{del2020PAB}, as shown in Fig.~\ref{fig:fit}. We thereby further reduce the sensitivity to noise in our experiment and hence the stability of our optimization protocol.

\subsection{Convergence of the minimal eigenvalue}

\begin{figure}[h]
    \centering
    \includegraphics[width =  \columnwidth]{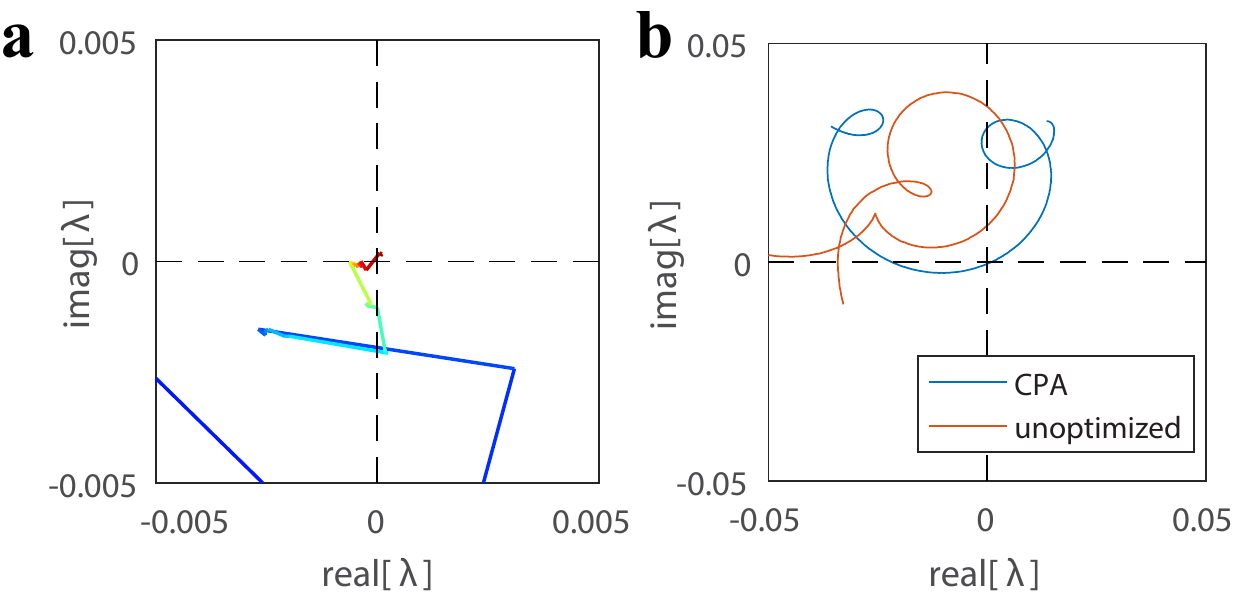}
    \caption{Minimal eigenvalue $\lambda$ of the scattering matrix shown in the complex plane as a function of the iterations for $\nu=\nu_{CPA}$ (a) and as a function of frequency for an optimized and an unoptimized configuration (b). In both cases, $\lambda \rightarrow 0$ when the CPA condition is obtained. }
    \label{fig:eigenvalue}
\end{figure}

Fig.~\ref{fig:eigenvalue} provides an illustration based on experimental data of how the smallest eigenvalue of the scattering matrix approaches zero, both at the targeted operation frequency as a function of iterations, and as a function of frequency for an optimal and unoptimized configuration.

\subsection{Characterization of the Chaotic Cavity}

\begin{figure}
    \centering
    \includegraphics[width =  \columnwidth]{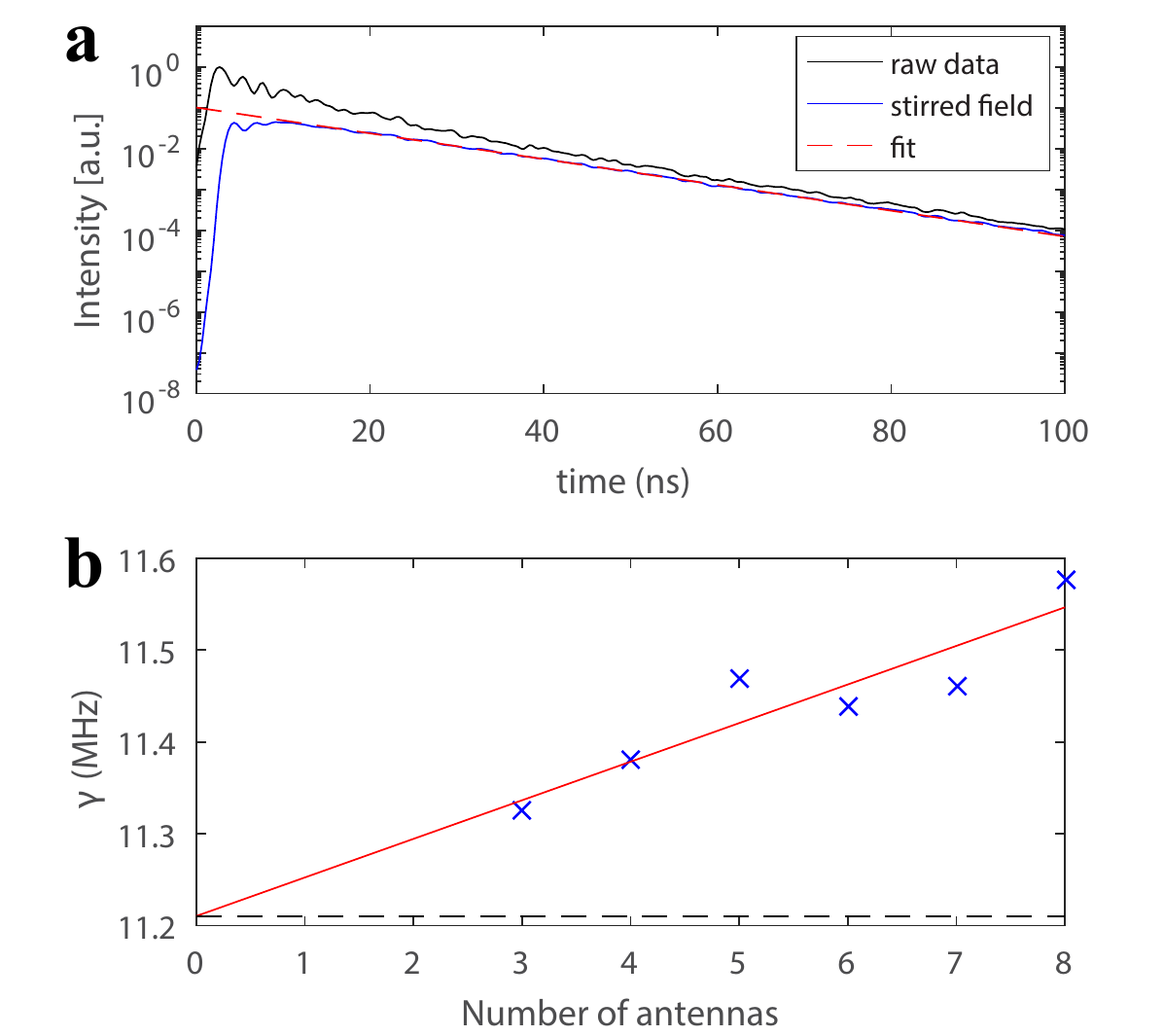}
    \caption{(a) Variations of the reflected intensity in the time domain averaged over 300 random configurations of the metasurfaces. The blue (black) curve is found from an inverse Fourier Transform of the scattering matrix elements with (without) removing its average value. The two curves converge at late times to the same exponential decay, $e^{-2\gamma t}$ (red dashed line). (b) By fitting the exponential decay for a number of channels to the cavity varying from 3 to 8, we extract $gamma$ as a function of $N$ (blue crosses). The red line is a linear regression of the data and the black dashed line gives the extracted absorption strength $\gamma_a$ using Eq.~(\ref{eq:gamma_fit}).}
    \label{fig:gamma_ant}
\end{figure} 

Here, we estimate the linewidth associated with the antennas in the chaotic process within the cavity. We first measure $S$ for 300 random configurations of the metasurfaces. In our experiment, the metasurfaces only partially cover the cavity walls and a substantial unstirred field components remains that is not impacted by the metasurface configuration. The ratio of unstirred to stirred field components $|\langle S_{ba}(f) \rangle|^2 / \langle |S_{ba}(f)|^2 \rangle $ is equal to 0.65 at $f=5.147$~GHz. Therefore, we remove the unstirred components $\langle S \rangle$ before estimating the average linewidth associated with the ports in the chaotic process. We then perform an inverse Fourier transform and find the intensity in the time domain $I(t)$. The average signal is presented in Fig.~\ref{fig:gamma_ant}a. We compare this signal with the intensity found in absence of the unstirred components, i.e. found from the field to which its average has been subtracted, ${S} = S - \langle S \rangle$. In both case, the intensity is seen to decrease exponentially, $I(t) \sim \text{exp}(-\Gamma t)$ for $t>25$~ns. We repeat this procedure for different numbers of antennas $N_{\text{ant}}$ connected to the cavity, varying $N_{\text{ant}}$ from 3 to 8. In each case, we extract $\gamma = \Gamma/2\pi$ which is shown in Fig.~\ref{fig:gamma_ant}b. This makes it possible to separate the contribution of the antennas $\gamma_n = N_{\text{ant}} \gamma_c$ ($\gamma_c$ is the average linewidth for a single antenna) and the contribution of internal losses within the cavity $\gamma_a$ as $\gamma$ can be decomposed as 
\begin{equation}
    \gamma = N_{\text{ant}} \gamma_c + \gamma_a.
\label{eq:gamma_fit}    
\end{equation}

From a linear fit of $\gamma$, we find that $N \langle \gamma_n \rangle = 0.33$~MHz ($N=8$), and $\langle \gamma_a \rangle = 11.2$~MHz. This shows that in the chaotic process, internal losses strongly exceeds losses through the antennas. This gives a quality factor of $Q = f/\gamma  =446 $ at $f = 5.147$~GHz. Finally, we now evaluate the modal overlap, $d = \gamma / \Delta$, of the cavity. The mean level spacing of successive resonances can be estimated from Weyl's law~\cite{weyl1911asymptotische,arendt2009weyl} as $\Delta = c_0^3/(8\pi V f^2) \sim 0.9$~MHz. We obtain $d = 12.8$, showing that many resonances contribute to the scattering matrix at any given frequency.

\begin{figure}[h]
    \centering
    \includegraphics[width =  \columnwidth]{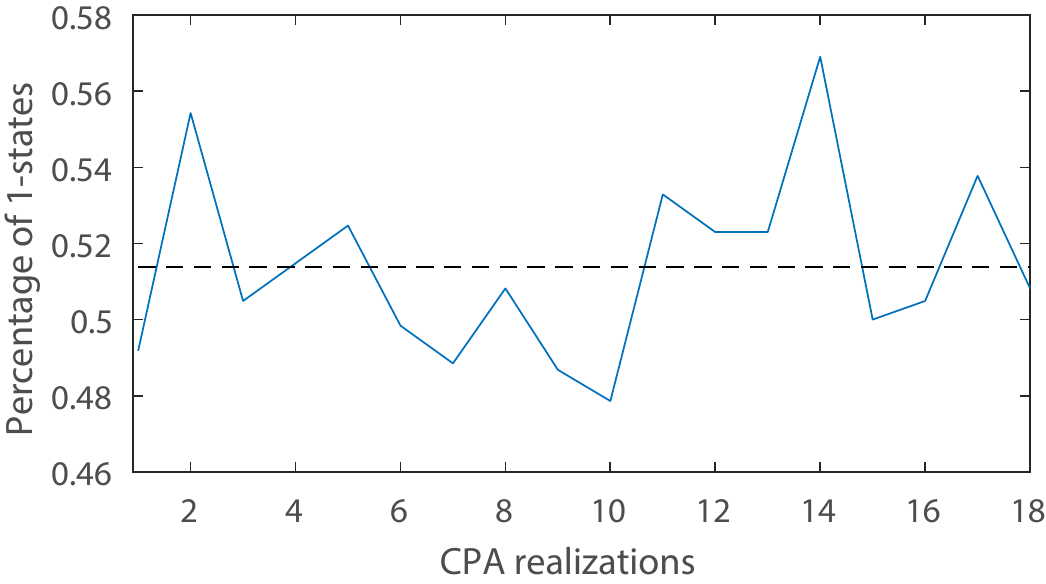}
    \caption{Percentage of meta-atoms in state ``1'' for 18 independent realizations of CPA.}
    \label{fig:percentage}
\end{figure}

\subsection{Analysis of Optimal Metasurface Configurations}

Here, we evaluate the percentage of meta-atoms configured to state ``1'' for each of our 18 independent CPA realizations. As seen in Fig.~\ref{fig:percentage}, this percentage is very close to 50~\% for all 18 realizations. Hence, as expected given the chaotic nature of our enclosure, neither state ``0'' nor state ``1'' dominates in the configurations corresponding to a CPA condition.

\bibliographystyle{apsrev4-1}

\providecommand{\noopsort}[1]{}\providecommand{\singleletter}[1]{#1}%

\end{document}